\documentclass[12pt,a4paper]{iopart}
\usepackage{graphicx}
\usepackage{cite}
\usepackage{color}
\usepackage{multirow}
\usepackage{epsfig}
\usepackage{epstopdf}
\usepackage{amssymb}
\usepackage{units}

\newcommand{\CO}{CO$_2$}
\newcommand{\DL}{D_\mathrm{L}}
\newcommand{\DT}{D_\mathrm{T}}
\newcommand{\DLnull}{D_{\mathrm{L}}^{(0)}}
\newcommand{\DTnull}{D_{\mathrm{T}}^{(0)}}

\newcommand{\nyeff}{\nu_{\mathrm{eff}}}
\newcommand{\tnyeff}{\tilde{\nu}_\mathrm{eff}}

\usepackage[deletedmarkup=xout]{changes}
\definechangesauthor[color=orange]{MV}
\definechangesauthor[color=magenta]{IK}
\definechangesauthor[color=violet]{DL}
\definechangesauthor[color=red]{NP}
\definechangesauthor[color=blue]{ZD}

\begin{document}

\title[Electron transport in \CO]{Electron transport parameters in \CO: scanning drift tube measurements and kinetic 
computations
}

\author{M. Vass$^1$, I. Korolov$^1$, D. Loffhagen$^2$, N. Pinh\~ao$^3$, Z. Donk\'o$^1$}

\address{$^1$ Institute for Solid State Physics and Optics, Wigner Research Centre for Physics,
Hungarian Academy of Sciences, 1121 Budapest, Konkoly Thege Mikl\'os str. 29-33, Hungary}
\address{$^2$ Leibniz Institute for Plasma Science and Technology (INP Greifswald), Felix-Hausdorff-Str. 2, 17489 Greifswald, Germany}
\address{$^3$ Instituto de Plasmas e Fus\~{a}o Nuclear, Instituto Superior T\'ecnico, Universidade de Lisboa, Av. Rovisco Pais, 1049-001 Lisboa, Portugal}

\ead{donko.zoltan@wigner.mta.hu}

\begin{abstract}
This work presents transport coefficients of electrons (bulk drift velocity,  longitudinal diffusion coefficient, and  effective ionization frequency) in \CO\/ measured under time-of-flight conditions over a wide range of the reduced electric field, $\unit[15]{Td} \leq E/N \leq \unit[2660]{Td}$ in a scanning drift tube apparatus. The data obtained in the experiments are also applied to determine the effective steady-state  Townsend ionization coefficient. These parameters are compared to the results of previous experimental studies, as well as to results of various  kinetic computations: solutions of the  electron  Boltzmann equation under different approximations (multiterm and density gradient expansions) and 
 Monte Carlo simulations.  The experimental data extend the range of $E/N$ compared with previous  measurements and are consistent with most of the transport parameters  obtained in these earlier studies. The computational results point out the range of applicability of the respective approaches to determine the  different measured transport properties of electrons in CO$_2$. They demonstrate as well the need for further 
 improvement of the electron collision cross section data for 
 CO$_2$ taking into account the present experimental data. 
\end{abstract}


\submitto{\PSST}
\maketitle
\section{Introduction}
\label{sec:Introduction}
The current (2016) atmospheric \CO\/ concentration is 
\unit[404.21]{ppm} and it undergoes a constant growth~\cite{lap}. In order to avoid its serious consequences foreseen at higher concentrations, the \CO\/ production must be reduced. One possible solution is the utilization of \CO\/, that is, the conversion of \CO\/ into more valuable chemical compounds that can be used as fuel or feedstock gas for different chemical processes (e.g.\  methanol or CO~\cite{barton,zhang,AeSoBo2015CSC702}). 

The amount of energy necessary for this conversion could be gained from renewable energy sources (e.g.\ wind and solar cells) during the periods when the production of electricity exceeds the demands. As mentioned above, \CO\/ can be split to CO an O$_2$ via the reaction 
$\mathrm{CO_2} \rightarrow \mathrm{CO} + \frac{1}{2} \mathrm{O_2}$. Plasma technologies have been gaining increasing interest for \CO\/ conversion ~\cite{ism,Annemie}.  The energy efficiency and the conversion rate have been examined in dielectric barrier discharges \cite{danhua,AeSoBo2015CSC702,BrWeSaEn2014JAP123303,yang,zheng,zhou}, corona discharges \cite{malik,morova,yuezhong}, gliding arc discharges~\cite{nunnaly}, radio-frequency discharges~\cite{laura}, and nanosecond repetitively pulsed discharges \cite{scapinello}. 

In order to promote the plasma-based chemical technologies, it is crucial to improve knowledge about the fundamental properties of the interactions taking place in the plasma phase, which are characterized by the  collision cross sections or rates of relevant plasma-chemical processes and transport parameters of relevant particles. Among all the particles, electrons play a central role. Therefore, their transport coefficients are of fundamental interest.

Drift tubes have been serving as the principal sources of transport coefficient data throughout several decades. In these systems low density clouds or ``swarms'' of electrons are created, which propagate under the influence of an external electric field. Based on their operation principles drift tube experiments have three major types \cite{Robson91}:  
\begin{itemize}
\item{Pulsed Townsend (PT) settings consist of two plane-parallel electrodes. Electron swarms are usually initiated by fast UV light pulses that induce photoemission of electrons from the negatively biased electrode. Recordings are made of the time-dependent displacement current pulses.}
\item{Time-of-flight (TOF) settings employ as well pulsed electron sources and make use of the collection of particles that arrive at a detector, which can operate on the basis of different principles. In this case, the same transport coefficients can usually be determined as in PT settings.}
\item{Steady-state Townsend (SST) settings operate with continuous electron sources and provide information about the Townsend ionization coefficient, via, e.g.\ the increase of the electron current with increasing electrode separation.}
\end{itemize}

In pulsed systems in the hydrodynamic regime 
with the electric field in the $-z$~direction, 
the  theoretical spatio-temporal distribution of 
the electron density of a swarm generated at time $t=0$ and position $z=0$  can immediately be 
derived from the solution of the continuity equation. It is given by~\cite{BlFl1984AJP593} 
\begin{equation}
n_\mathrm{e}(z,t) = \frac{n_0}{(4 \pi \DL t)^{1/2}} \exp \left[ \nyeff t - \frac{(z - Wt)^2}{4 \DL t} \right] \, ,
\label{eq:n}
\end{equation}
where $n_0$ is the initial electron density, $\DL$ is the longitudinal diffusion coefficient, $\nyeff$ is the effective ionization frequency (equal to 
ionization frequency minus attachment frequency), and
$W$ is the {\it bulk drift velocity}, which gives the velocity of the center-of-mass of the electron cloud.

In PT experiments the measured displacement current is proportional to the spatial integral of $n_\mathrm{e}(z,t)$ under hydrodynamic conditions. 
In contrast, in TOF systems (including our experimental system) the measured signal is 
directly proportional to $n_\mathrm{e}(z,t)$ (under hydrodynamic conditions), and the extraction of the transport coefficients proceeds via fitting the measured signals with the theoretical form (\ref{eq:n}). This procedure yields the 
bulk drift velocity $W$, the longitudinal diffusion coefficient $\DL$, and effective ionization frequency $\nu_{\rm{eff}}$. 

Our experimental apparatus  
allows ``mapping'' of the electron swarms \cite{swarm1,jpd-paper} by taking measurements at a 
large 
number of drift lengths. The swarm maps generated this way allow a visual observation of the equilibration of the transport as well, and a straightforward distinction between hydrodynamic and non-hydrodynamic domains can be done.  Besides the ``basic'' TOF transport coefficients ($W$, $\DL$, and $\nyeff$), 
the application of the well-known connection between TOF and SST transport coefficients also allows us to derive the 
effective 
(steady-state) Townsend ionization 
coefficient~$\alpha$~\cite{TaSaSa1977JPDAP1051,BlFl1984AJP593}. 
 
In the present paper we revisit the electron transport coefficients  in carbon dioxide in the range of reduced electric fields between 
 15 and \unit[2660]{Td}, where $\unit[1]{Td} = \unit[10^{-21}]{V m^{-2}}$. 
 Our measured data are compared with experimental results  of earlier studies compiled in the review of Dutton~\cite{Du1975JPCRD577} as well 
 as of more recent analyses reported 
 in~\cite{Chachereau,Hernandez,Yoshinaga,Hasegawa,Elford}.  Corresponding theoretical studies of electron transport properties in \CO\/ have 
 been presented e.g.\  in~\cite{DeXi2014JJAP096201,YoUrJuBaHe2009IEEETPS764,
 BrWiWi1986INCD681,BrWiWi1985LNC365,KueLu1979JPDAP2123,HaPh1967PR70}. 
 These results were obtained by Boltzmann equation 
and/or Monte Carlo simulation methods, where different data for the
electron collision cross sections were used.
 
In addition to the experimental investigations we also carry out computations of the transport coefficients using different types of theoretical methods. Besides Monte Carlo simulations, the electron Boltzmann equation is solved under different assumptions and approximations. The application of these different approaches allows us to mutually verify the accuracy of the different methods, test the assumptions used by each method and uncover errors in the codes~\cite{Pinhao04}. Furthermore, information about the respective transport properties provided by each method is given.  The numerical calculations and simulations are performed using the recently published set of electron collision cross sections 
 reported in~\cite{Grofulovic16}. 
  
The manuscript is organized as follows. 
In section \ref{sec:ExpSet} we give a concise description of our 
experimental setup and introduce the methods of data evaluation. A  discussion of the various computational methods and the resulting transport properties is presented in 
section \ref{sec:NumMeth}. It is followed by the discussion of the results in section~\ref{sec:Res}. This section comprises the presentation of the present experimental results 
and their comparison with previously available 
measured data in section~\ref{sec:Expdata} as well as the 
comparison between transport parameters computed using the various numerical methods 
and the present experimental data in section~\ref{sec:Modellingresults}. 
Section~\ref{sec:Summary} gives our concluding remarks. 

\section{Experimental apparatus and data acquisition}
\label{sec:ExpSet}

\subsection{Experimental setup}

The experiments are based on a ``scanning'' drift tube apparatus of which the details have been presented in \cite{swarm1} and which has already been applied for the measurements of transport coefficients of electrons in various gases: argon, synthetic air, methane, and deuterium \cite{jpd-paper}. Our system -- in contrast with previously developed drift tubes -- allows recording of ``swarm maps'' that show the spatio-temporal development of electron clouds under TOF conditions. The simplified scheme of our experimental apparatus is shown in figure~\ref{fig:expsetup} and its brief description is given below. 

\begin{figure}[ht]
\begin{center}
\includegraphics[width =0.6\textwidth]{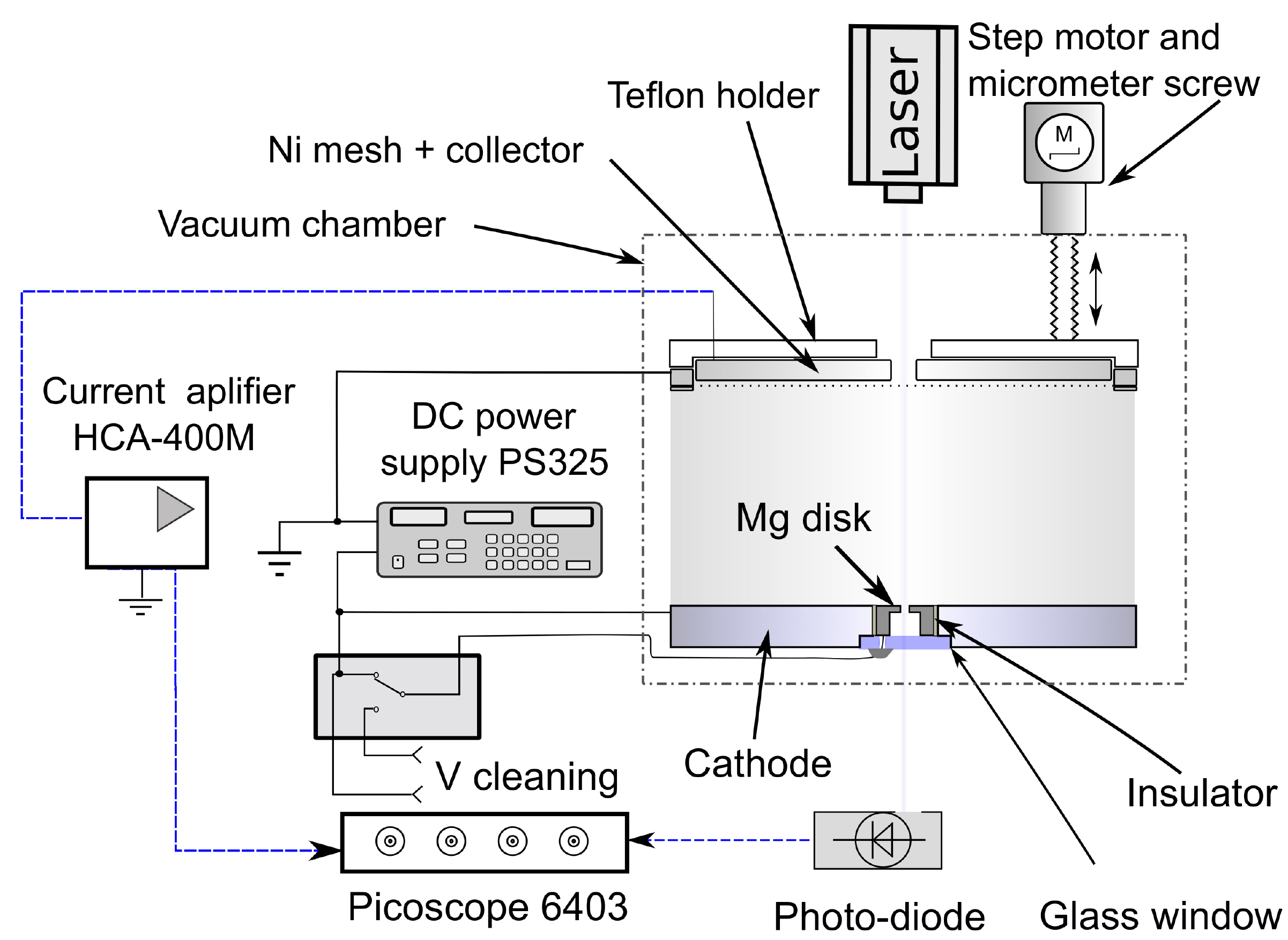}
\caption{Simplified scheme of the experimental setup. (Copyright IOP Publishing Ltd. Reproduced with permission from \cite{jpd-paper}.)}
\label{fig:expsetup}
\end{center}
\end{figure}

Electron swarms are initiated at \unit[3]{kHz} repetition rate by 
$\unit[1.7]{\mu J}$, $\approx \unit[6]{ns}$ pulses of a frequency-quadrupled diode-pumped YAG laser that irradiates a magnesium disk embedded inside a stainless steel cathode electrode having a diameter of 105 mm (see figure~\ref{fig:expsetup}). The swarms move under the influence of an electric field applied between the cathode and a grounded nickel grid 
(with \unit[88]{\%}  transmission and 
45 lines/inch density) situated at \unit[1]{mm} (fixed) distance in front of a stainless steel collector electrode. The grid and the collector are moved together by a step motor connected to a micrometer screw mounted via a vacuum feedthrough to the vacuum chamber that encloses the drift cell. The distance between the cathode and the collector is scanned in \unit[1]{mm} steps over the range of $L = 13.6 - \unit[63.6]{mm}$. The electric field is kept constant during the scanning process by automatically adjusting the cathode--collector voltage by a PS-325 (Stanford Research Systems) power supply. The laser pulses are also used for triggering the data collection. The current generated by the electrons entering the grid-collector gap is amplified by a high speed current amplifier (type Femto HCA-400M) and is acquired by a digital oscilloscope (type Picoscope 6403B) with sub-ns time resolution. The measured current signal is proportional to the flux of the electrons entering the grid-collector gap. 
As the flux and the density of the electrons are proportional in the hydrodynamic regime,   the data shown in the form of ``swarm maps'' may also be interpreted as the electron density.
We note that the non-hydrodynamic behavior can directly be identified in the swarm maps as it was demonstrated earlier \cite{swarm1}. This (i) allows us to restrict our data evaluation procedures to the domain where hydrodynamic equilibrium prevails, and (ii) makes it unnecessary to find this domain by numerical simulations of the experimental system. 

Preceding the experiments the vacuum chamber is evacuated by a turbomolecular pump backed by a rotary pump down to a base pressure of $\unit[10^{-5}]{Pa}$ for several days. During the experiments a slow (\unit[5]{sccm}) flow of (5.0 purity) CO$_2$ gas is established by a flow controller, and the gas pressure inside the chamber is measured by a Pfeiffer CMR~362 capacitive gauge. The experiment is fully controlled by a LabView program. The measurements are performed at room temperature of \unit[293]{K}. In order to prevent the electrical breakdown of the gas the experiments had to be conducted at different pressures, i.e., at different values of $N$, for different domains of the whole $E/N$ range. Of course, this was done with an overlap of the $E/N$ values when changing the pressure, to allow observation of an eventual dependence of the experimental data on the change of pressure. As we used the hydrodynamic regime for the evaluation of the swarm characteristics such dependence was actually not observed.

\subsection{Data acquisition}

As it was already mentioned in 
section {\ref{sec:Introduction}},  the transport coefficients $W$, $D_{\rm L}$, and $\nu_{\rm eff}$ are derived by fitting the experimentally measured 
signal of the current $I(z,t)$  to the theoretical form (\ref{eq:n}) 
of the electron density $n_\mathrm{e}(z,t)$ describing the 
spatio-temporal evolution of the swarm. It is important that this fitting is executed for hydrodynamic conditions. Deviations from the hydrodynamic condition are easy to recognize in the measured swarm maps, however \cite{swarm1}. In cases when such deviations are observed, the fitting is executed for a part of the space-time domain where equilibrium transport prevails.
In this fitting procedure we select the complete map, as well as its sub-domains, and accept the resulting values of the transport coefficients only when the fits over different domains give data within a pre-defined deviation from each other (see below).

The uncertainties of the determination of the transport coefficients mainly originate from the errors and fluctuations of the experimental characteristics: errors of the setting of the electrode gap, the  pressure, the voltage across the cell, the (fluctuating) laser intensity, and electric noise. The fitting procedure was found not to enhance the uncertainties and fluctuations in the determination of $W$, as in this case the {\it position} of the current peak is the most important quantity that is well defined even in the presence of noises. This is why we estimate the total accuracy of the determination of $W$ to be of the order of 3\% based on our estimations of the errors in the measurements of the voltage, pressure, etc. The uncertainty of $W$ caused by the fitting procedure itself (selection of the fitting domain) proved to be less than 1\%. On the other hand, the values of the resulting $D_{\rm L}$ were sensitive on the choice of the domain of fitting. Here, we accept values that are within 15\%, which seems to be a proper value  considering the fact that previous studies have also reported $D_{\rm L}$ values with such an accuracy. The larger sensitivity of the measured $D_{\rm L}$ on the choice of the fitting domain originates from the fact that the determination of the width of the pulse is more sensitive on the signal to noise ratio, compared to the determination of the peak position.

Besides the fitting of the theoretical and measured density distributions, we also apply a slicing method for the determination of the bulk drift velocity $W$ for $E/N$ values below 1000 Td: cutting the $I(z,t)$ maps of the swarm (or the $n_\mathrm{e}(z,t)$ density distribution (\ref{eq:n})) at  fixed values of time,  symmetrical Gaussian functions are obtained and the peaks of these functions can be associated with the center-of-mass of the swarm. A straight line fit to this position as a function of time yields the value of $W$ \cite{swarm1}. The results obtained for $W$ by both the fitting procedure and the slicing method agree more closely than \unit[1]{\%}.

Having determined the TOF transport coefficients $W$, $\DL$, and $\nyeff$, the effective Townsend ionization coefficient, $\alpha$, characteristic of 
SST experiments, is calculated according to 
\begin{equation}
\frac{1}{\alpha} = \frac{W}{2 \nu_{\rm eff}} + \sqrt{\left( \frac{W}{2 \nu_{\rm eff}} \right)^2 - \frac{\DL}{\nu_{\rm eff}}} \, ,
\label{eq:conn}
\end{equation}
based on the discussions in the papers of Tagashira \textit{et al.}~\cite{TaSaSa1977JPDAP1051} and of Blevin and Fletcher~\cite{BlFl1984AJP593}. In the absence of diffusion, i.e., $\DL=0$,  eq.~(\ref{eq:conn}) reduces to $\alpha = \nu_{\rm eff}/ W$. This value is increased in the presence of diffusion and, for the gases and conditions covered here, this increase is between \unit[1]{\%}
 and \unit[20]{\%}.

\section{Numerical methods}
\label{sec:NumMeth}

The experimental studies of the electron transport parameters are supplemented by results obtained by numerical modeling and simulation. In addition to Monte Carlo (MC) simulations, three different methods are applied to solve the Boltzmann equation (BE)  for  electron swarms in a background gas with density $N$ and acted upon by a constant electric field, $\vec{E}$, and assuming hydrodynamic conditions: 
(i) a multiterm method for the solution of the time- and space-independent Boltzmann equation ,
(ii) a multiterm approach for solving the spatially one-dimensional, steady-state electron Boltzmann equation as well as (iii) the $S_n$ method applied to a density gradient expansion of the electron distribution function. They differ in their initial physical assumptions and in the numerical algorithms used and provide different properties of the electrons. In the following, a brief description of these three methods as well as main aspects of the MC simulation approach are given. In this discussion, the electric field is along the $z$ axis pointing in negative direction, $\vec{E}=-E\vec{e}_z$, and $\theta$ is the angle between $\vec{v}$ and $\vec{E}$. Moreover, we assume that after a sufficiently long relaxation time and length the  transport properties of the electrons do not change with time~$t$ and distance $z$ any longer, i.e., the electrons have reached a hydrodynamic regime characterizing a state of equilibrium of the system where the effects of collisions and forces are dominant and the electron velocity distribution function $f(\vec{r},\vec{v},t)$ has lost any memory of the initial state. 

\begin{figure}[!ht]
\begin{center}
\includegraphics[width=0.5\textwidth]{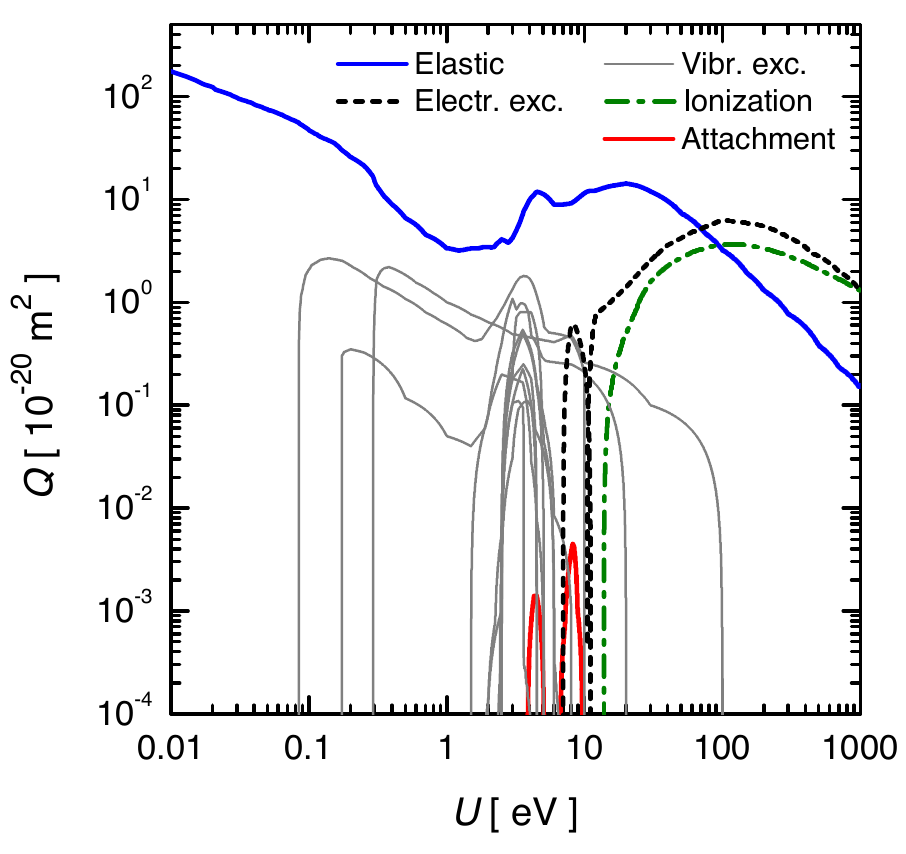} 
\caption{\CO\/ cross sections used in the kinetic computations. For more detailed information (specifications, threshold energies, etc.) about the individual processes see~\cite{Grofulovic16}.}
\label{fig:cs1}
\end{center}
\end{figure}

For our studies we use the electron--\CO\/ collision cross section set recently published in~\cite{Grofulovic16}, however, we neglect superelastic collision processes. The data set used includes the momentum transfer cross section for elastic collisions,  11 vibrational and two electronic excitation cross sections, the total 
 electron-impact ionization cross section and the collision cross section for dissociative electron attachment to \CO\/.  The cross section data are illustrated 
 in figure \ref{fig:cs1}. 

\subsection{Multiterm method for spatially homogeneous conditions}
\label{sec:MTMhomogenous}
To study the electron movement under the conditions mentioned above, we need further assumptions on the system and the electron velocity distribution function. In our first Boltzmann equation approach (abbreviated by BE~0D in the figures shown in section~\ref{sec:Res}) we consider a spatially homogeneous system where the electron density changes exponentially 
in time according to $n_\mathrm{e}(t) \propto \exp(\nyeff t)$, depending on the effective ionization frequency $\nyeff$. In this case we can neglect the  dependence of $f$ on the space coordinates and write the velocity distribution function under hydrodynamic conditions as
\begin{equation}
f(\vec{v},t) = \hat{f}(\vec{v}) n_\mathrm{e}(t) \, .
\label{eq:f_hat}
\end{equation}
The corresponding microscopic and macroscopic properties of the electrons are determined by the time-independent, spatially homogeneous Boltzmann equation for $\hat{f}(\vec{v})$. This distribution is symmetric around the field direction, $\hat{f}(\vec{v}) = \hat{f}(v, v_z/v)$, where $v_z$ is the $z$ component of the velocity $\vec{v}$ with magnitude~$v$. Thus, an expansion of the velocity distribution function with respect to $v_z/v \equiv \cos \theta$ in Legendre polynomials $P_n(\cos \theta)$ according to 
\begin{equation}
\hat{f}(v, \cos \theta) = \frac{1}{2 \pi} \left( \frac{m_\mathrm{e}}{2}\right)^{3/2} 
\sum_{n=0}^{l-1} \tilde{f}_n(U) P_n(\cos \theta)
\label{eq:LegendrePolynomialExpansion}
\end{equation}  
becomes possible, where the magnitude of the velocity was replaced by the kinetic energy $U = m_\mathrm{e} v^2/2$ of the electrons with mass $m_\mathrm{e}$ on the right-hand side. Substitution of expansion~(\ref{eq:LegendrePolynomialExpansion}) including an arbitrary number $l$ of expansion coefficients into the electron Boltzmann equation finally leads to a hierarchy of partial differential equations for the expansion coefficients~$\tilde{f}_n(U)$ with $n=0, \ldots, l-1$.  The normalization condition is $\int_0^\infty \tilde{f}_0(U) U^{1/2} \mathrm{d}U = 1$.
The resulting set of equations with typically eight expansion coefficients is solved employing a generalized version of the multiterm solution technique for weakly ionized steady-state plasmas~\cite{LeLoWi1998CPC33} adapted to take into account ionizing and attaching electron collision processes. The need for such multiterm approximation for the analysis of steady-state plasmas was found to arise in general when the lumped collision cross section of all inelastic 
collision processes is large and becomes comparable with the total cross section for elastic collisions over large parts of the relevant energy regions\cite{BrWiWi1985LNC365,WiBrWi1985ICPIG22}. Under such conditions, the coupling in the hierarchy of expansion coefficients $\tilde{f}_n$ leads to a large anisotropy of the electron velocity distribution function.

The macroscopic properties for spatially homogeneous conditions, namely the \textit{flux} drift velocity
\begin{equation}
w = - \mu E \, ,
\label{eq:fluxdriftvelocity}
\end{equation}
where $\mu$ is the mobility, the \textit{flux} diffusion coefficients~$\DLnull$ and~$\DTnull$,  the ionization and attachment frequencies~$\nu_{\mathrm{i}}$ and $\nu_{\mathrm{a}}$, can be obtained using the equations given in~\cite{WiWiBr1986INCD641,GrBeLo2009PRE036405} 
\begin{eqnarray}
\mu N & = & - \frac{e_0}{3} \left( \frac{2}{m_\mathrm{e}}\right)^{1/2} 
\int_0^{\infty} 
\frac{1}{Q_\mathrm{eff}(U)}\times \nonumber \\ 
& & \quad \left[ U \left(
\frac{\partial}{\partial U} \tilde{f}_0(U) 
+ 
\frac{2}{5}
\frac{\partial}{\partial U} \tilde{f}_2(U)
\right) 
+ 
\frac{3}{5} \tilde{f}_2(U)
\right] \mathrm{d}U 
\label{eq:MT1delectronmobility}  \, , \\
\DLnull N & = & 
\frac{1}{3} 
  \left( \frac{2}{m_\mathrm{e}}\right)^{1/2} 
\int_0^{\infty} 
\frac{U}{Q_\mathrm{eff}(U)} 
\left( \tilde{f}_0(U) + 
\frac{2}{5} \tilde{f}_2(U) \right) \mathrm{d}U
\label{eq:MT1delectronlongitudinaldiffusioncoefficient} \, , \\
\DTnull N& = & 
\frac{1}{3} 
  \left( \frac{2}{m_\mathrm{e}}\right)^{1/2} 
\int_0^{\infty} 
\frac{U}{Q_\mathrm{eff}(U)} 
\left( \tilde{f}_0(U) - \frac{1}{5} \tilde{f}_2(U) \right) \mathrm{d}U
\label{eq:MT1delectrontransversaldiffusioncoefficient} \, , \\
\nu_{\mathrm{i,a}}/N &=& 
  \left( \frac{2}{m_\mathrm{e}}\right)^{1/2} 
\int_0^{\infty} U Q^{\mathrm{i,a}}(U) \tilde{f}_0(U)
\mathrm{d}U \, .
 \label{eq:0dnyiandnya} 
\end{eqnarray}
Here,  $e_0$ is the elementary charge, 
\begin{equation}
Q_\mathrm{eff}(U) = Q^\mathrm{T}(U)+(2U/m_e)^{-1/2}\nyeff/N, 
\end{equation}
with $Q^\mathrm{T}$ being the sum of the elastic momentum 
transfer cross section and all inelastic cross sections, 
$Q^{\mathrm{i}}$ and $Q^{\mathrm{a}}$ denote 
the ionization and attachment cross sections, and 
$\nyeff = 
\nu_\mathrm{i} -\nu_\mathrm{a}$.
Furthermore, the effective  
ionization coefficient for spatially homogeneous plasmas is given by 
\begin{equation}
\alpha^{(0)} =
\frac{\nyeff}{w}  \, . 
  \label{eq:alphafrom0D}
\end{equation}

\subsection{Multiterm method for spatially inhomogeneous conditions}
\label{sec:MTMinhomogenous}

In the second Boltzmann equation approach (designated as BE~1D~SST below) we consider the spatial relaxation of an electron swarm. In this case we determine the effective Townsend ionization coefficient, $\alpha$, characteristic of SST experiments. Here, an idealized SST experiment with plane-parallel geometry similar to~\cite{WiLoSi2002ASS50,DuWhPe2008JPDAP245205} is considered, where a steady flux of electrons is emitted from the cathode. These electrons are accelerated in the positive $z$ direction under the action of the  electric field, ionize the gas or are attached by it. At a sufficiently large distance $z$ from the cathode, the mean transport properties of the electrons do not vary with position any longer and the electron density $n_\mathrm{e}$ assumes the exponential dependence on the distance 
\begin{equation}
n_\mathrm{e}(z) = c \exp{\left(\alpha z\right)} \, ,
\label{eq:neexponentialform}
\end{equation} 
where $c$ is a constant. 

In order to determine the microscopic and macroscopic properties of the electrons under such conditions, the spatially one-dimensional Boltzmann equation for their velocity distribution function~$f(\vec{r}, \vec{v})$ is solved. 
As the electric field and the inhomogeneity in the plasma are both parallel to the $z$ axis, this distribution function is also symmetric around the field direction, i.e., $f(\vec{r}, \vec{v}) = f(z, v, \cos \theta)$. 
Thus, we can expand the velocity distribution function 
in Legendre polynomials $P_n(\cos \theta)$ in accordance 
with (\ref{eq:LegendrePolynomialExpansion}), where the expansion 
coefficients $f_n$ are functions of $z$ and $U$ now 
and the normalization on the electron density according to 
\begin{equation}
n_\mathrm{e}(z) = \int_0^{\infty} U^{1/2} f_0(z,U) \mathrm{d}U \, 
\label{eq:1dElectrondensity} 
\end{equation}
holds. Substitution of this expansion into the  Boltzmann equation of the electrons results in a set of partial differential equations for the expansion coefficients~$f_n(z,U)$ with $n = 0, \ldots, l-1$ in the end. 
This equation system is solved numerically in accordance with the multiterm solution method described in
\cite{Lo2016IOPBookChapter3} using typically $l=8$ expansion coefficients. 

  The consistent particle balance of the electrons reads  
  \begin{equation}
  \frac{\mathrm{d}\phantom{z}}{\mathrm{d} z} \left(n_\mathrm{e}(z) v_\mathrm{m}(z) \right) = 
  n_\mathrm{e}(z) \nu_{\mathrm{i}}(z) - n_\mathrm{e}(z) \nu_{\mathrm{a}}(z) \, , 
  \label{eq:1dElectronParticlebalance}
\end{equation}  
where appropriate energy space averaging over the normalized expansion coefficients $\tilde{f}_n(z,U) = f_n(z,U)/n_\mathrm{e}(z)$ with $n=0$ and~1 yields the space-dependent  mean electron velocity~$v_\mathrm{m}$ in $z$ direction as well as the  space-dependent  frequencies of ionization $\nu_{\mathrm{i}}$  and of attachment $\nu_{\mathrm{a}}$, respectively. The latter are determined according to~(\ref{eq:0dnyiandnya}). The mean velocity given as~\cite{GrBeLo2009PRE036405}
 \begin{equation}
v_\mathrm{m}(z) = \frac{1}{3} \left( \frac{2}{m_\mathrm{e}}\right)^{1/2} 
 \int_0^{\infty} U \tilde{f}_1(z,U)
 \mathrm{d}U \label{eq:1dmeanvelocity} 
 \end{equation}
is composed of the space-dependent \textit{flux} drift velocity 
$w(z) = - \mu(z) E$, which is  
in accordance with (\ref{eq:fluxdriftvelocity}), 
and a diffusion part. 
It can be written
as 
\begin{equation} 
v_\mathrm{m}(z) = w(z) 
- \frac{1}{n_\mathrm{e}(z)} 
\frac{\mathrm{d}}{\mathrm{d} z} 
\left( n_\mathrm{e}(z) \DLnull(z) \right) \, ,
\label{eq:vmdriftdiffusion}
\end{equation}
where the space-dependent electron mobility~$\mu(z)$ and the 
\textit{flux} longitudinal 
diffusion coefficient~$\DLnull(z)$ as well as the  \textit{flux} transverse diffusion coefficient~$\DTnull(z)$ are given by similar integration over $\tilde{f}_n(z,U)$
as~(\ref{eq:MT1delectronmobility}), (\ref{eq:MT1delectronlongitudinaldiffusioncoefficient}) and 
(\ref{eq:MT1delectrontransversaldiffusioncoefficient}) 
with $Q_\mathrm{eff}$ replaced by $Q^\mathrm{T}$~\cite{GrBeLo2009PRE036405,WiWiBr1986INCD641}.

When approaching SST conditions at a sufficiently large distance from the cathode, the normalized expansion coefficients $\tilde{f}_n$ and, consequently, the macroscopic properties $v_\mathrm{m}$, $\mu$, $\DLnull$, $\DTnull$, and  $\nyeff = \nu_{\mathrm{i}} - \nu_{\mathrm{a}}$ become independent of the position~$z$. Then, the particle balance 
equation~(\ref{eq:1dElectronParticlebalance}) becomes 
\begin{eqnarray}
v_\mathrm{m}^\mathrm{(S)} 
\frac{\mathrm{d} n_\mathrm{e}(z) }{\mathrm{d} z} 
  & = &  
w^\mathrm{(S)}
\frac{\mathrm{d} n_\mathrm{e}(z) }{\mathrm{d} z} 
- 
\DL^\mathrm{(S)}
\frac{\mathrm{d}^2 n_\mathrm{e}(z) }{\mathrm{d} z^2} 
=
  n_\mathrm{e}(z) 
  \nyeff^\mathrm{(S)} \, ,
  \label{eq:SSTElectronParticlebalance} 
  \end{eqnarray}
  where the upper index $\mathrm{(S)}$ denotes the SST condition.
Using equations~(\ref{eq:neexponentialform}) 
and~(\ref{eq:SSTElectronParticlebalance}), thus the effective 
Townsend ionization coefficient is directly given by 
\begin{eqnarray}
\alpha  = 
\frac{\nyeff^\mathrm{(S)}}{v_\mathrm{m}^\mathrm{(S)}}   
  \label{eq:SSTalphafrom1Dnyeff}
\end{eqnarray}
or dependent on the non-observable quantities $w^\mathrm{(S)}$,  
$\DL^\mathrm{(S)}$, and 
$\nyeff^\mathrm{(S)}$~\cite{BlFl1984AJP593} according to 
\begin{eqnarray}
 \alpha    =  
  \left\{
  \frac{w^\mathrm{(S)}}{2 \nyeff^\mathrm{(S)}}
  + 
  \left[ \left( \frac{w^\mathrm{(S)}}{2 \nyeff^\mathrm{(S)}} \right)^2 
  - \left( \frac{\DL^\mathrm{(S)}}{\nyeff^\mathrm{(S)}} \right) \right]^{1/2}  
  \right\}^{-1} \, . 
  \label{eq:SSTalphafrom1Drelationinverse}
\end{eqnarray}
Notice that although relation~(\ref{eq:SSTalphafrom1Dnyeff}) has the same form as the effective 
ionization coefficient~(\ref{eq:alphafrom0D}) for spatially homogeneous conditions, 
both parameters are generally not identical. Furthermore, the effective (steady-state) Townsend ionization coefficient calculated from the TOF transport coefficients according to~(\ref{eq:conn}) looks similarly but gives a different result from relation~(\ref{eq:SSTalphafrom1Drelationinverse}). However, the values of $\alpha$ obtained from~(\ref{eq:conn}) and   (\ref{eq:SSTalphafrom1Drelationinverse}) are identical, if higher-order terms in the derivation of~(\ref{eq:conn}) are neglected, as usually assumed in the TOF case~\cite{TaSaSa1977JPDAP1051,BlFl1984AJP593}.

Following Blevin and Fletcher~\cite{BlFl1984AJP593}, 
the SST parameters can be expressed by a 
series expansion with respect to $\alpha$. 
The application of such expansion and its 
correlation with the density gradient expansion 
discussed in section~\ref{sec:DGM} make it possible 
to approximately determine the 
\textit{bulk} drift velocity~$W$ 
from the SST calculations. 
Then, we can get $W$ from the approximation equation 
\begin{equation}
W \approx w - v_\mathrm{m}^\mathrm{(S)} + 
\frac{\nyeff}{\alpha} = W_{\mathrm{a}} \, ,
\label{eq:WfromSSTand0dresults}
\end{equation} 
provided that the \textit{flux} drift velocity $w$ 
according to (\ref{eq:fluxdriftvelocity})  and the effective ionization  
frequency~$\nyeff$  
for spatially homogeneous 
conditions are known 
in addition to the mean velocity~$v_\mathrm{m}^\mathrm{(S)}$ 
and 
the effective Townsend ionization coefficient  $\alpha$ 
at SST conditions. 

Finally, it should be mentioned that the application of 
(\ref{eq:neexponentialform}) and the assumption of SST conditions 
also 
leads to a set of equations for $\tilde{f}^\mathrm{(S)}_n(U)$. 
This equation system can also be solved efficiently by a modified  
version of the multiterm technique method~\cite{LeLoWi1998CPC33} 
adapted to treat SST conditions. Corresponding results show  
excellent agreement with the SST results obtained by using the method 
BE~1D~SST and are therefore not included in the figures presenting our results, in favour of clarity. 

\subsection{Density gradient representation}
\label{sec:DGM}

The third Boltzmann equation approach to describe the electron swarm at hydrodynamic conditions 
(labelled as BE~DG~TOF below) is based on an expansion of the electron velocity distribution function, 
$f$, on the consecutive space gradients of the electron density $n_\mathrm{e}$.
In this case, $f$ depends on $(\vec{r},t)$ only through the density $n_\mathrm{e}(\vec{r},t)$
and can be written as an expansion on the gradient operator $\nabla$ according to
\begin{equation}
f(\vec{r},\vec{v},t) = \sum_{j=0}^\infty
F^{(j)}(\vec{v})\stackrel{j}\odot(-\nabla)^{j}n_e(\vec{r},t) \, .
\end{equation}
Here, the expansion coefficients $F^{(j)}(\vec{v})$ are tensors of order $j$ depending only on $\vec{v}$, and $\stackrel{j}\odot$ indicates a $j$-fold scalar product~\cite{Kumar80}. Note that the first coefficient $F^{(0)}(\vec{v})$ corresponds to the conventional distribution function for  homogeneous conditions, denoted by $\hat{f}(\vec{v})$ in equation (\ref{eq:f_hat}) in section~\ref{sec:MTMhomogenous}. 

Each expansion coefficients $F^{(j)}$ of order $j$ is obtained from a hierarchy of equations for each component depending on the previous orders and all with the same structure.  In particular, a total of five equations is required,  namely for the coefficients components $F^{(0)}$, $F^{(1)}_z$, $F^{(1)}_\mathrm{T}$, $F^{(2)}_{zz}$ and  $F^{(2)}_\mathrm{TT}$.
 In the present code these equations are solved using a variant of the finite element method given in \cite{SEGUR1983116} in a $(v,\cos\theta)$ grid.

The transport parameters for a TOF experiment are obtained from the above expansion coefficients as 
\begin{eqnarray}
W &=& 
  \int v_z F^{(0)}(\vec{v}) \mathrm{d}\vec{v} 
+ \int \tnyeff(v)
F_z^{(1)}(\vec{v}) \mathrm{d} \vec{v}, \label{eq:DG_W} \\ 
\DL &=& \int v_z F_z^{(1)}(\vec{v}) \mathrm{d}\vec{v} + 
\int \tnyeff(v) F_{zz}^{(2)}(\vec{v}) \mathrm{d}\vec{v} 
\label{eq:DG_DL}, \\
\DT &=& \frac{1}{2}\left\{\int v_\mathrm{T} 
F_\mathrm{T}^{(1)}(\vec{v}) \mathrm{d}\vec{v} + \int
\tnyeff(v) F_\mathrm{TT}^{(2)}(\vec{v}) 
\mathrm{d}\vec{v}\right\}, \label{eq:DG_DT} \\
\nyeff &=&  \int \tnyeff(v)
F^{(0)}(\vec{v}) \mathrm{d} \vec{v}, \label{eq:DG_nu} 
\end{eqnarray}
where $W$, $\DL$, $\DT$, and $\nyeff$  are  the 
\textit{bulk} drift velocity, the 
\textit{bulk} diagonal 
longitudinal and transverse 
components of the diffusion tensor, and 
the effective ionization frequency, respectively, and  
$\tnyeff(v) = v N [Q^\mathrm{i}(v) - Q^\mathrm{a}(v)]$. 
Here, the lower indexes $z$ and $\mathrm{T}$ on the right indicate the longitudinal and transverse components of the vectors $\vec{v}$ and $F^{(1)}$ and the diagonal terms of the tensor $F^{(2)}$.
The corresponding \textit{flux} components of the drift velocity and the diffusion tensor are the first terms on the right-hand side of equations (\ref{eq:DG_W})--(\ref{eq:DG_DT}). The second term on the  right-hand side of these equations describes the explicit contribution of the non-conservative collision processes by the velocity-space averaging over the product of $\tnyeff(v)$ and the expansion coefficients of order 1 and 2, respectively. 

The effective or apparent Townsend ionization coefficient $\alpha$, as determined in SST experiments, can be computed either from $F^\mathrm{(S)}(\vec{v})$, the distribution function 
at SST conditions obtained from the solution of an additional equation similar to the one for the expansion coefficient $F^{(0)}$, as~\cite{Yousfi85} 
\begin{equation}
\alpha = \frac{\int \tnyeff(v)
F^\mathrm{(S)}(\vec{v}) \mathrm{d} \vec{v}}{\int v_z
F^\mathrm{(S)}(\vec{v}) \mathrm{d} \vec{v}}
\end{equation}
or from the TOF parameters using the relation~\cite{TaSaSa1977JPDAP1051,BlFl1984AJP593} 
\begin{equation}
\alpha = \frac{W}{2 \DL} 
- \sqrt{\left(\frac{W}{2 \DL}\right)^2 - 
\frac{\nyeff}{\DL}  } \, ,
\label{eq:alphaTfromDGM}
\end{equation}
which is the inverse representation of equation~(\ref{eq:conn}).

It can be shown that the \textit{flux} component ($w$) of (\ref{eq:DG_W}) has the same expression as (\ref{eq:1dmeanvelocity}) and that 
equation (\ref{eq:DG_nu}) divided by $N$ is equivalent to 
$(\nu_\mathrm{i} - \nu_\mathrm{a})/N$ as given by (\ref{eq:0dnyiandnya}). 
In the case of the diffusion tensor, the comparison is more complex.  Using the Boltzmann equation for each component of $F^{(1)}$, we can rewrite equations (\ref{eq:DG_DL}) and (\ref{eq:DG_DT}) 
as~\cite{Yousfi85,Pinhao04}  
\begin{eqnarray}
\DL &=& \int
\frac{v_z-W}
{\tilde{\nu}_\mathrm{T}(v)+\nyeff}  
v_z F^{(0)}(\vec{v}) \mathrm{d} \vec{v} - 
\vec{a} \int \frac{v_z}
{\tilde{\nu}_\mathrm{T}(v)+\nyeff} 
\nabla_v F_z^{(1)}(\vec{v})
\mathrm{d} \vec{v} \nonumber \\ && 
+ 
\int \tnyeff(v) F_{zz}^{(2)}(\vec{v}) 
\mathrm{d}\vec{v}\,, \label{eq:DG_DL2} \\ 
\DT &=& \frac{1}{2}\left\{ \int
\frac{v_\mathrm{T}^2}
{\tilde{\nu}_\mathrm{T}(v)+\nyeff}  
F^{(0)}(\vec{v}) \mathrm{d} \vec{v}
- \vec{a} \int
\frac{v_\mathrm{T}}
{\tilde{\nu}_\mathrm{T}(v)+\nyeff}  
\nabla_v 
F_\mathrm{T}^{(1)}(\vec{v}) \mathrm{d}\vec{v} 
\right. \nonumber \\ && \left.
+ \int
\tnyeff(v) F_\mathrm{TT}^{(2)}(\vec{v})
\mathrm{d}\vec{v}\right\}\,, \label{eq:DG_DT2}
\end{eqnarray}
where $\vec{a}$ is the acceleration due to the electric field and $\tilde{\nu}_\mathrm{T}(v) = v N Q^\mathrm{T}(v)$. 
However, only the terms involving $F^{(0)}$ are comparable with equations~(\ref{eq:MT1delectronlongitudinaldiffusioncoefficient}) and (\ref{eq:MT1delectrontransversaldiffusioncoefficient}). The second terms represent contributions from the electric field and the gradient of  the distribution of electrons with different velocities inside the swarm, as given by $F^{(1)}$, to the spreading of the swarm. The third terms are  the contribution of non-conservative processes.
Using the index~''$(F^{(0)})$'' to designate the first term  of (\ref{eq:DG_DL2}) and (\ref{eq:DG_DT2}), after an expansion in Legendre polynomials and a change of variables from $v$ to $U$, these terms can be written as
\begin{eqnarray}
\DL^{(F^{(0)})}N &=& 
\frac{1}{3}\left(\frac{2}{m_\mathrm{e}}\right)^{1/2}
\int_0^\infty \frac{U}{Q_\mathrm{eff}(U)}
\left(\tilde{f}_0(U)  
 + \frac{2}{5}\tilde{f}_2(U) \right)
\mathrm{d}U 
\nonumber \\ && 
- \frac{W}{3} \int_0^\infty 
\frac{U^{1/2}}{Q_\mathrm{eff}(U)} \tilde{f}_1(U) 
\mathrm{d}U \, ,
\label{eq:DLF0} \\ 
\DT^{(F^{(0)})}N &=& 
\frac{1}{3}
\left(
\frac{2}{m_\mathrm{e}} 
\right)^{1/2}
\int_0^\infty \frac{U}{Q_\mathrm{eff}(U)}
\left(\tilde{f}_0(U)-\frac{1}{5}\tilde{f}_2(U)\right)
\mathrm{d}U \, . 
\label{eq:DTF0}
\end{eqnarray}
Except for the last term in (\ref{eq:DLF0}), these are the same expressions as (\ref{eq:MT1delectronlongitudinaldiffusioncoefficient}) and  (\ref{eq:MT1delectrontransversaldiffusioncoefficient}).
Notice that a similar representation 
was derived in~\cite{WiWiBr1986INCD641} for the case of conservative 
electron collision processes. 

\subsection{Monte Carlo method}
\label{sec:MC}

In the MC simulation technique we trace the trajectories of the electrons 
in the external electric field and under the influence of collisions. Due to the low degree of ionization under the swarm conditions considered here, only electron-background gas molecule collisions are taken into account. The motion of the electrons between collisions is described by their equation of motion 
\begin{equation}
\label{eq:mot}
m_\mathrm{e} \frac{\mathrm{d}^2 \vec{r} }{\mathrm{d}t^2} 
= -e_0 \vec{E} .
\end{equation}
The determination of the electron trajectories between collisions is carried out by integrating (\ref{eq:mot}) numerically over 
time steps of duration $\Delta t$ ranging between 
0.5 and $\unit[2.5]{ps}$
for the various conditions. While this procedure is totally deterministic, the collisions are handled in a stochastic manner. The probability of the occurrence of a collision is computed after each time step, for each of the electrons, as 
\begin{equation}
P(\Delta t)=1-\exp\left[-N v~ Q^{\mathrm{T}}(v) \Delta t\right].
\label{eq:Collisionprobability}
\end{equation}
Comparison of $P(\Delta t)$ with a random number $R_{01}$ having a uniform distribution over the $[0,1)$ interval allows deciding about the occurrence of a collision: if $R_{01} \leq P(\Delta t)$ a collision is simulated. This is carried out in the center-of-mass frame (for a more detailed description see, e.g., \cite{MCDonko}). The type of collision is also determined in a random manner; the probability $P_k$ of process $k$ at a given energy $U$ is given by 
\begin{equation} 
P_k=\frac{Q_k(U)}{Q^{\mathrm{T}}(U)} ,
\end{equation} 
where $Q_k(U)$ is the collision cross section of the $k$-th process. 
The elastic collisions are assumed to result in isotropic scattering. Accordingly, we use the elastic momentum transfer cross section.

We carry
 out two types of MC simulations:  
\begin{itemize}
\item TOF simulations (labeled as MC~TOF below) are executed 
to determine the bulk ($W$) and flux ($w$) drift velocities of the electrons according to 
\begin{equation}
W = \frac{\mathrm{d}\phantom{t}}{\mathrm{d}t} \left[ \frac {\sum_{j=1}^{N_\mathrm{ e}(t)} z_j(t) }{N_\mathrm{e}(t)} \right]
\label{eq:bulk}
\end{equation}
and
\begin{equation}
w = \frac{1}{N_\mathrm{e}(t)} \sum_{j=1}^{N_\mathrm{ e}(t)} \frac{\mathrm{d}z_j(t)}{\mathrm{d}t}.
\label{eq:flux}
\end{equation}
Here,  $N_\mathrm{e}(t)$ is the number of electrons in the swarm at time $t$. 
Note that when expanding (\ref{eq:bulk}), 
we obtain the relation between these velocities as 
$W = w - \langle z \rangle \, \nyeff $, where $\langle z \rangle$ is the average position and 
\begin{equation}
\nyeff = \frac{\mathrm{d}(\ln N_e(t))} {\mathrm{d}t},
\label{eq:nueffTOFMC}
\end{equation}
showing how $W$ includes a contribution from non-conservative processes. These simulations also yield the longitudinal and transversal diffusion coefficients $\DL$ and $\DT$ obtained as \cite{Hu1977AJP83} 
\begin{eqnarray} 
\DL & = & \frac{1}{2} \frac{\mathrm{d} [\langle z^2(t)\rangle-\langle z(t)\rangle ^2 ]}{\mathrm{d} t} \, ,
\label{eq:DLTOFbyMC} \\
\DT & = & \frac{1}{4} \frac{\mathrm{d} [\langle x^2(t) + y^2(t) \rangle ]}{\mathrm{d} t} \,  .
\end{eqnarray}
Furthermore, the effective Townsend ionization coefficient $\alpha$ can be calculated according to relation (\ref{eq:conn}) using (\ref{eq:bulk}), (\ref{eq:nueffTOFMC}), 
and (\ref{eq:DLTOFbyMC}) 
and neglecting 
higher-order terms in the derivation of~(\ref{eq:conn})  
as usually done in the TOF case~\cite{TaSaSa1977JPDAP1051,BlFl1984AJP593}. 
\item SST simulations (named as MC~SST below) are additionally used to derive directly the 
effective Townsend ionization coefficient $\alpha$ from the spatial growth of the electron density under stationary conditions according to 
\begin{equation}
\alpha = \frac{1}{n_\mathrm{e}(z)}\frac{{\rm d}n_\mathrm{e}(z)}{{\rm d}z} \,  .
\end{equation} 
\end{itemize}

\subsection{Summary of computational methods}

The computational methods used to derive the transport coefficients of electrons for different conditions, as well as the resulting transport coefficients are summarized in Table~\ref{tab:methods}.

\begin{table}[htb]
\centering
\caption{Computational methods and their identifiers, as well as the transport coefficients obtained from the specific methods in this work. Recall the $W$, $D_{\rm L}$ and $\nu_{\rm eff}$ are obtained in the experiments and the ``experimental'' $\alpha$ is obtained according to eq.~(\ref{eq:conn}) from these coefficients. 
Notice that $W_\mathrm{a}$ in the row denoted by 
BE~1D~SST is calculated according 
to~(\ref{eq:WfromSSTand0dresults})
using results of method BE~0D as well. 
\label{tab:methods}}
    \begin{footnotesize}
    
\begin{tabular}{llllllll}
\br
Id. & Method & 
\multicolumn{3}{l}{Transport coefficients}
\\
 & & bulk & flux & $n_\mathrm{e}$ change \\
\mr
BE 0D & Spatially homogeneous BE &  &  $w$, $\DLnull$, $\DTnull$ & $\nyeff$, $\alpha^{(0)}$\\
BE 1D SST & 1-dimensional stationary BE & $W_\mathrm{a}$
 & $w^\mathrm{(S)}$, $\DL^\mathrm{(S)}$, $\DT^\mathrm{(S)}$ & $\nyeff^\mathrm{(S)}$, $\alpha$\\
BE DG TOF & Density gradient expansion of BE & 
$W$, $D_{\rm L}$, $D_{\rm T}$ 
& $w$ &   $\nu_{\rm eff}$, $\alpha$\\
BE DG SST & Density gradient expansion of BE &  &  & 
\phantom{$\nu_{\rm eff}$,} 
 $\alpha$\\ 
MC TOF & Time-of-flight MC simulation & $W$, $\DL$, $\DT$  & $w$  
& $\nu_{\rm eff}$, $\alpha$\\
MC SST & Stationary MC simulation & & & 
\phantom{$\nu_{\rm eff}$,}  $\alpha$\\
\br
\end{tabular}

\end{footnotesize}
\end{table} 

\section{Results}
\label{sec:Res}

Measurements of the electron transport coefficients 
have been performed in the wide range of the reduced 
electric field between 15 and \unit[2660]{Td} 
at a gas temperature~$T$ of \unit[293]{K}. 
The results of our measurements are presented and discussed 
in the following.  In section~\ref{sec:Expdata}, they are compared 
with previous experimental data. 
Section~\ref{sec:Modellingresults} adds a comprehensive 
comparison with results obtained by the different Boltzmann equation 
methods and by MC simulations. 

\subsection{Experimental data and comparison with previous experimental results}
\label{sec:Expdata}

\begin{figure}[t!]
\begin{center}
\includegraphics[width=0.8\textwidth]{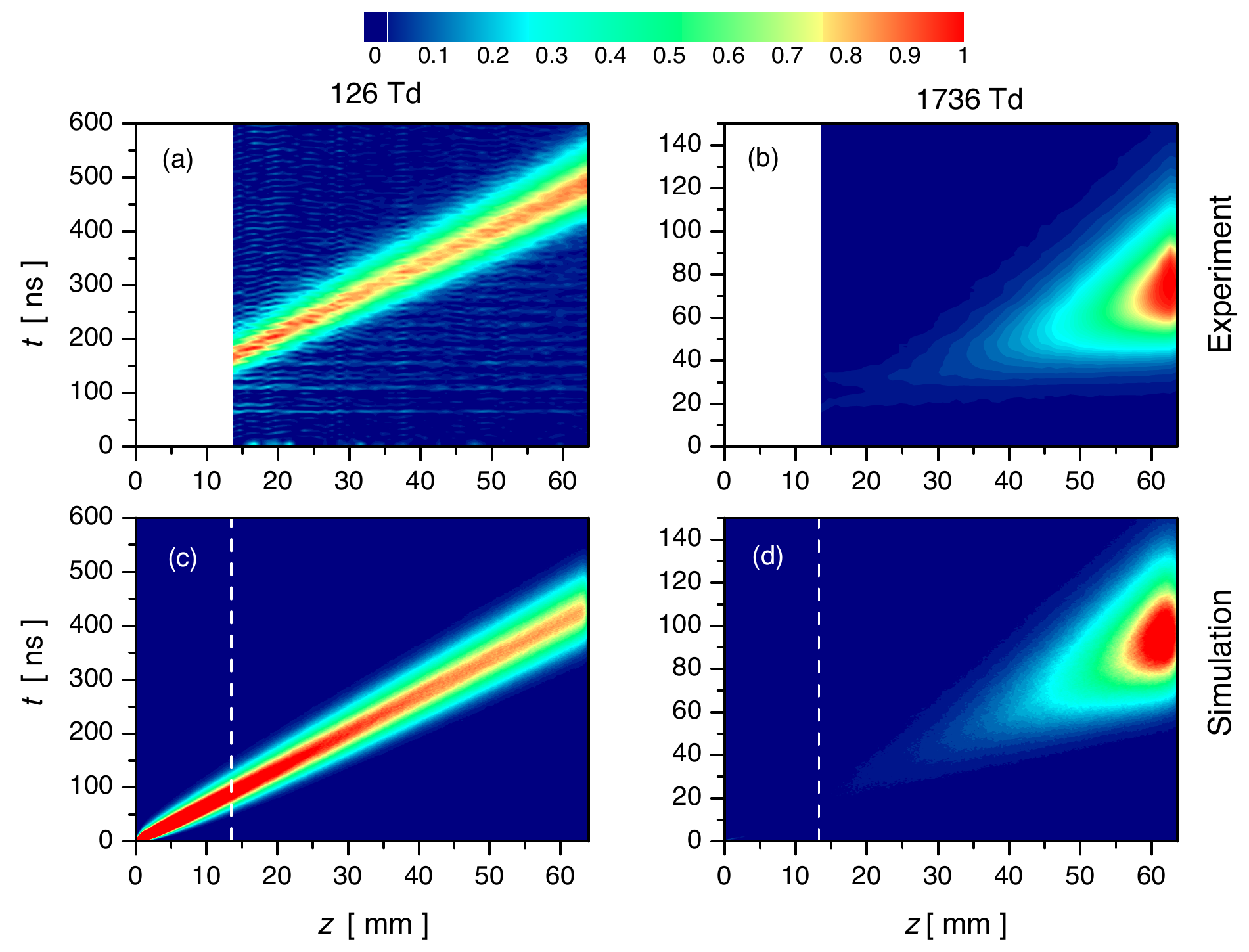} 
\caption{Maps of electron swarms in CO$_2$ at $E/N = \unit[126]{Td}$ (left column) and \unit[1736]{Td} (right column). The top panels show experimental data and the bottom panels display corresponding  data obtained by MC simulation. The vertical dashed white lines in the bottom row of panels indicate the minimum electrode gap of \unit[13.6]{mm} in the experiment (cf.~figure~\ref{fig:expsetup}).}
\label{fig:maps}
\end{center} 
\end{figure}

We begin the presentation of the results by illustrating the 
measured swarm maps in comparison with simulated maps 
obtained by the MC~TOF method. 
Figure~\ref{fig:maps} shows the maps for two different sets of experimental 
conditions with $E=\unit[6942]{V m^{-1}}$ and $N=\unit[5.51 \times 10^{22}]{m^{-3}}$ as well as $E=\unit[5347]{V m^{-1}}$ and $N=\unit[3.08 \times 10^{21}]{m^{-3}}$
at room temperature, corresponding to the values of the reduced electric field $E/N = \unit[126]{Td}$ and \unit[1736]{Td}, respectively. The maps allow a straightforward visual observation of the characteristic effects of drift, diffusion and ionization. In this representation of the spatio-temporal distribution of the electron density the inclination of the path of the electron cloud corresponds to its drift. The widening of the cloud is related to the effect of diffusion, while the increase of density with the spatial coordinate and with time is a signature of avalanching through ionization processes at these ionization-dominated conditions. Notice that for attachment-dominated conditions with $E/N$ between about 30 and \unit[90]{Td}~\cite{Hernandez} a decrease of the density should equally be observable. At the low reduced electric field of $E/N = \unit[126]{Td}$  (figures~\ref{fig:maps}(a) and (c)) the electron cloud exhibits a small spreading and small density increase, indicating low rate of diffusion and low effective ionization. This behavior changes remarkably when $E/N$ is increased. At \unit[1736]{Td} the increasing importance of both diffusion and ionization processes changes the character of the swarm map noticeably as illustrated in figures~\ref{fig:maps}(b) and (d). These features are well represented by both the experimental and simulated results.   

\begin{figure}[ht]
\begin{center}
\includegraphics[width=0.45\textwidth]{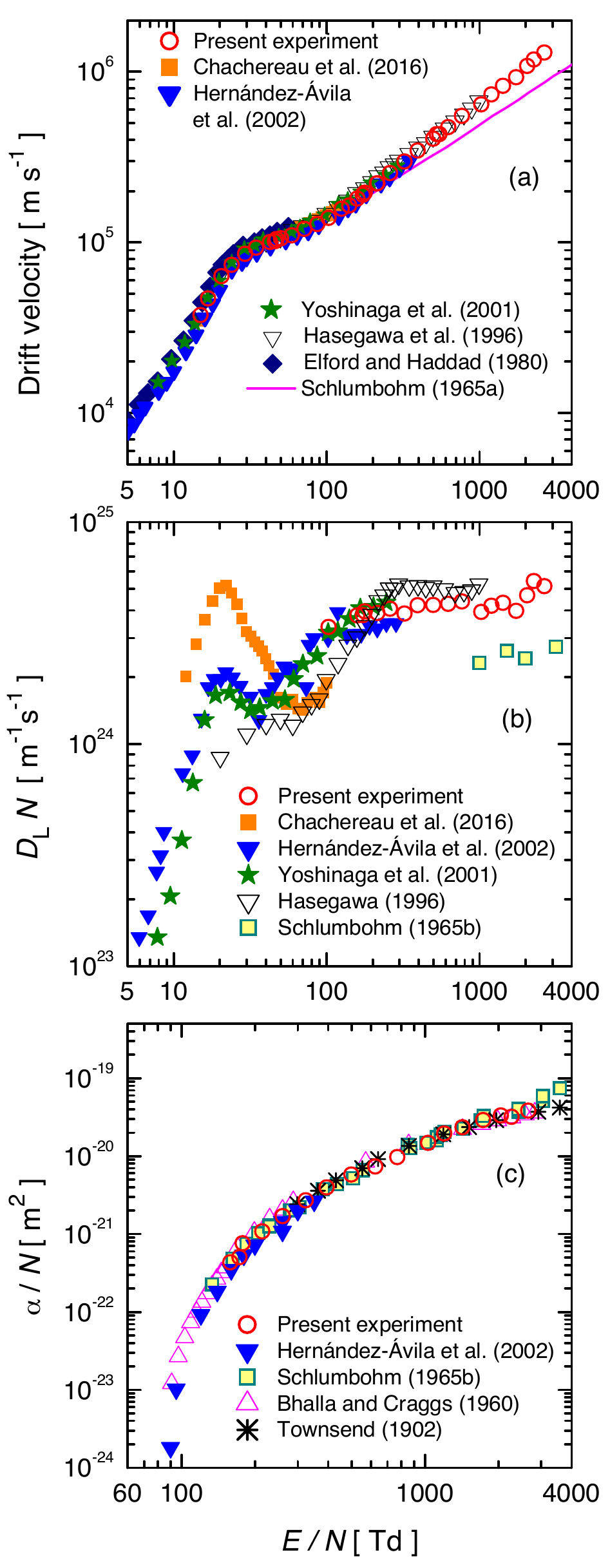} 
\caption{Transport coefficients derived from our 
experiments (a) $W$, (b) $D_{\rm L} N$, and (c) $\alpha / N$ 
in comparison with 
results of previous studies: 
Chachereau {\it et al.} \cite{Chachereau}, 
Hern\'andez-\'Avila {\it et al.} \cite{Hernandez}, 
Yoshinaga {\it et al.} \cite{Yoshinaga}, 
Hasegawa {\it et al.} \cite{Hasegawa}, 
 Elford and Haddad \cite{Elford}, 
Schlumbohm (1965a) \cite{SchlumbohmDrift}, 
Schlumbohm (1965b) \cite{SchlumbohmDL}, 
Bhalla and Craggs \cite{Bhalla}, 
and Townsend \cite{Townsend}.}
\label{fig:exp-and-comp}
\end{center}
\end{figure}

Our experimental results for the transport coefficients are displayed in figure \ref{fig:exp-and-comp}. The panels (a), (b), and (c) show the measured bulk drift velocity $W$, the measured  
longitudinal diffusion coefficient times gas number density $\DL N$ and the reduced effective Townsend ionization coefficient $\alpha/N$, respectively. The latter is calculated according to (\ref{eq:conn}) using the measured effective ionization frequency $\nyeff$ in addition to $W$ and $\DL$. Tabulated values of the transport coefficients as a function of the reduced electric field 
are also given in tables~\ref{tableW} and~\ref{tableW2}. Table~\ref{tableW} contains the measured values of the bulk drift velocity at $E/N$ between 15 and \unit[2660]{Td}. Table~\ref{tableW2} gives the longitudinal diffusion coefficient, effective ionization frequency, and the effective Townsend ionization coefficient in the 
range of the reduced electric field from 159 to \unit[2660]{Td}.

\Table{\label{tableW} 
Measured bulk drift velocity $W$ of electrons in CO$_2$  at \unit[293]{K}.}
\br

$E/N$ & $W$ & $E/N$&$W$ &  $E/N$ & $W$ & $E/N$ & $W$\\
$[\mathrm{Td}]$ & $[\unit[10^4]{m~s^{-1}}]$ & $[\mathrm{Td}]$ & $[\unit[10^4]{m~~~s^{-1}}]$ & $[\mathrm{Td}]$ & $[\unit[10^4]{m~~ s^{-1}}]$ &
$[\mathrm{Td}]$ & $[\unit[10^4]{m~s^{-1}}]$ \\
\mr

15.0 &	3.75 &	50.4&	10.5 & 173 &	19.0 &  625&	47.3\\
16.8 &	4.68 &	59.4&	11.0 & 178 &	19.6 &  769&	55.1\\
20.6 &	6.35 &	71.8&	12.0 & 214 &	22.1 &  1030&	64.3\\
24.0 &	7.39 &	87.1&	12.9 & 260 &	25.5 &  1200&	73.6\\
29.4 &	8.56 &	87.2&	12.8 & 324 &	29.7 &  1430&	82.6\\
34.6 &	9.31 &	103&	13.9 & 395 &	34.6 &  1740&	92.8\\
42.8 &	10.1 &	126&	15.7 & 499 &	40.5 &  2050&	108\\
46.0 &	10.1 &	141&	16.4 & 525 &	43.0 &  2270&	118\\
47.2 &	10.5 &	159&	18.1 & 547 &	43.0 &  2660&	130\\	

\br
\end{tabular}
\end{indented}
\end{table}

\Table{\label{tableW2} 
Measured longitudinal diffusion coefficient times gas number density $\DL N$, measured 
reduced ionization frequency $\nyeff / N$ and 
reduced effective Townsend ionization coefficient $\alpha / N$ calculated using 
(\ref{eq:conn}) of electrons in CO$_2$  at \unit[293]{K}. }
\br

$E/N$ & $\DL N$ & $\nyeff/N$  & $\alpha/N$  \\
$[\mathrm{Td}]$  & $[\unit[10^{24}]{m^{-1}s^{-1}}]$ 
& $[ \unit[10^{-14}]{m^{3}s^{-1}}]$ 
& $[\unit[10^{-20}]{m^{2}}]$   \\
\mr

159  &	3.79  &  0.00777 &	0.0432 \\  
173  &	3.81  &  0.00950 &	0.0504 \\  
178  &	3.96  &  0.0147  &	0.0759 \\  
214  &	3.89  &  0.0235  &	0.108  \\  
260  &	4.07  &  0.0421  &	0.169  \\  
324  &	3.87  &  0.0772  &	0.270  \\  
395  &	4.22  &  0.129   &	0.395  \\  
499  &	4.23  &  0.218   &	0.578  \\  
625  &	4.28  &  0.316   &	0.732  \\
769  &	4.39  &  0.475   &	0.968  \\
1030 &	3.94  &  0.833   &	1.48   \\
1200 &	4.19  &  1.27    &	1.95   \\
1430 &	4.32  &  1.71    &	2.35   \\
1740 &	3.98  &  2.31    &	2.89   \\
2050 &	4.68  &  3.06    &	3.31   \\
2270 &	5.42  &  3.28    &	3.19   \\
2660 &	5.14  &  4.19    &	3.83   \\

\br
\end{tabular}
\end{indented}
\end{table}

Our experimental results are compared with experimental data 
reported in~\cite{Chachereau,Hernandez,Yoshinaga,Hasegawa,%
Elford,SchlumbohmDrift,SchlumbohmDL,Bhalla,Townsend}. Specific features of these experimental investigations are given in the following. Chachereau {\it et al.}~\cite{Chachereau} used a 
PT apparatus with a back-illuminated photocathode. The electron drift velocity $W$ and the longitudinal electron diffusion coefficient $\DL$ were determined by fitting the measured displacement current waveform with a theoretical form. Furthermore, the effective ionization rate coefficient $(\nyeff/N)$ was obtained using an electron swarm model. Results for CO$_2$ were reported for reduced electric fields between 10 and \unit[100]{Td}.  Hern\'andez-\'Avila {\it et al.} \cite{Hernandez} used as well a PT system equipped with a nitrogen laser to initiate photoelectron pulses. Displacement current pulses were measured at a fixed gap length of \unit[30]{mm} at $E/N$ values ranging from 2 to \unit[350]{Td}. These studies provided results for $W$, $\DL$, and the effective ionization coefficient $\tilde{\alpha} = \nyeff/W$, which is identical to $\alpha$ according to~(\ref{eq:conn}) only in the absence of diffusion~\cite{Du1975JPCRD577}. Yoshinaga {\it et al.} \cite{Yoshinaga} employed a double-shutter drift tube with variable gap length between 10 and \unit[50]{mm}. Photoelectrons were initiated by pulsed UV light from a quartz window covered with a gold film. The bulk drift velocity and longitudinal diffusion coefficient were measured over the $E/N$ range from~8 to \unit[300]{Td} in this experiment. Hasegawa {\it et al.}  \cite{Hasegawa} determined the electron drift velocity as well as the ratio $\DL / \mu$ of the longitudinal diffusion coefficient to the electron mobility (longitudinal characteristic energy) from the arrival time spectra of electrons in a double-shutter drift tube. Using both these measured quantities, we computed the $\DL N$ values shown in figure~\ref{fig:exp-and-comp}(b). Their experiment covered the 
range of $\unit[20]{Td} \leq E/N \leq \unit[1000]{Td}$. 
Elford and Haddad~\cite{Elford} used the Bradbury-Nielsen type TOF method to measure the drift velocity of electrons at different temperatures. Their measurements extended over $E/N$ values between 0.1 and \unit[50]{Td} at $T = \unit[293]{K}$. 
Schlumbohm used 
a PT apparatus, operated with photoelectrons initiated by short 
($\approx \unit[15]{ns}$) UV light pulses and described 
in~\cite{Schlumbohm1965Method},  
to measure the drift velocity~\cite{SchlumbohmDrift} 
for about $\unit[450]{Td} \leq E/N \leq \unit[6060]{Td} $. These results were given in functional form of $W$ depending on $E/p$ at \unit[293]{K} and corresponding tabulated data can be found e.g.\ in~\cite{Du1975JPCRD577}. In addition, measured results of the effective ionization coefficient $(\tilde{\alpha})$ for about $ \unit[135]{Td} \leq E/N \leq \unit[4350]{Td} $ and 
 the ratio $\DL / \mu$ were reported in~\cite{SchlumbohmDL}. The latter was used to determine the longitudinal diffusion coefficient in the $E/N$ range from 1000 to \unit[3162]{Td}~\cite{Du1975JPCRD577}. The SST method was used in~\cite{Bhalla,Townsend} to determine 
 the effective Townsend ionization coefficient~$\alpha$.  Bhalla and Craggs~\cite{Bhalla} carried out measurements for the  $E/N$ range from about 90 to \unit[2825]{Td} and Townsend's data~\cite{Townsend} cover the range between about  300 and  \unit[3600]{Td}. Corresponding tabulated data can be found e.g.\ in~\cite{Du1975JPCRD577}. 
%
%
Our drift velocity data (figure~\ref{fig:exp-and-comp}(a)) agree well with most of those given in 
these 
previous works, which also show a good degree of consistency. The only exception concerns the experimental data of Schlumbohm~\cite{SchlumbohmDrift} at $E/N$ larger than \unit[450]{Td}, where increasing differences are obvious. The agreement between our data and the data given by Hasegawa {\it et al.}~\cite{Hasegawa} is particularly good and our measurements extend the range of $E/N$ from \unit[1000]{Td} in~\cite{Hasegawa} to 2660 Td. We estimate the experimental error of the present bulk drift velocity data to be less than \unit[3]{\%} below  \unit[1000]{Td} and less than \unit[5]{\%} above this value. 

The values of the longitudinal diffusion coefficient times gas number density $\DL N$ shown in 
figure~\ref{fig:exp-and-comp}(b) exhibit larger scattering, which is explained by the higher uncertainty of the determination of $\DL$ in the experiments  ($\approx \unit[15]{\%}$ including our present data) as compared to that of the drift velocity. Our data for $\DL N$ cover the range $\unit[159]{Td} \leq E/N \leq \unit[2660]{Td}$. Up to $E/N = \unit[1000]{Td}$, our data agree reasonably with those of Hasegawa {\it et al.}~\cite{Hasegawa} considering the uncertainties of both data sets quoted above. The $\DL N$ values derived by Schlumbohm \cite{SchlumbohmDL} for  $E/N \geq \unit[1000]{Td}$ are about a factor of two lower compared to the present data.

Figure \ref{fig:exp-and-comp}(c) shows our experimental results for the effective Townsend ionization coefficient $\alpha$ in comparison with the corresponding data provided 
by Bhalla and Craggs~\cite{Bhalla} and by Townsend~\cite{Townsend}  as well as the results for the effective ionization coefficient $\tilde{\alpha}$ of Hern\'andez-\'Avila {\it et al.}~\cite{Hernandez} and Schlumbohm~\cite{SchlumbohmDL}. The different data sets show 
generally very good agreement. Our data cover the range of $\unit[159]{Td} \leq E/N \leq \unit[2660]{Td}$, for which the estimated experimental error of the data is approximately 
$\unit[8]{\%}$. For $E/N < \unit[159]{Td}$ the accuracy of the determination of $\alpha$  decreased drastically in our experiments due to the worse signal to noise ratio 
so that no data is available here. 

\subsection{Numerical results for transport coefficients and comparison with experimental data}
\label{sec:Modellingresults}

In this section we present the results obtained by the various numerical methods described in section~\ref{sec:NumMeth} using the cross section set from \cite{Grofulovic16}. We compare the results with each other as well as with the present experimental data. A summary of the methods used to derive the transport coefficients of electrons for different conditions, as well as the resulting transport coefficients were given in table~\ref{tab:methods}.

\begin{figure}[!ht]
\begin{center}
\includegraphics[width=0.50\textwidth]{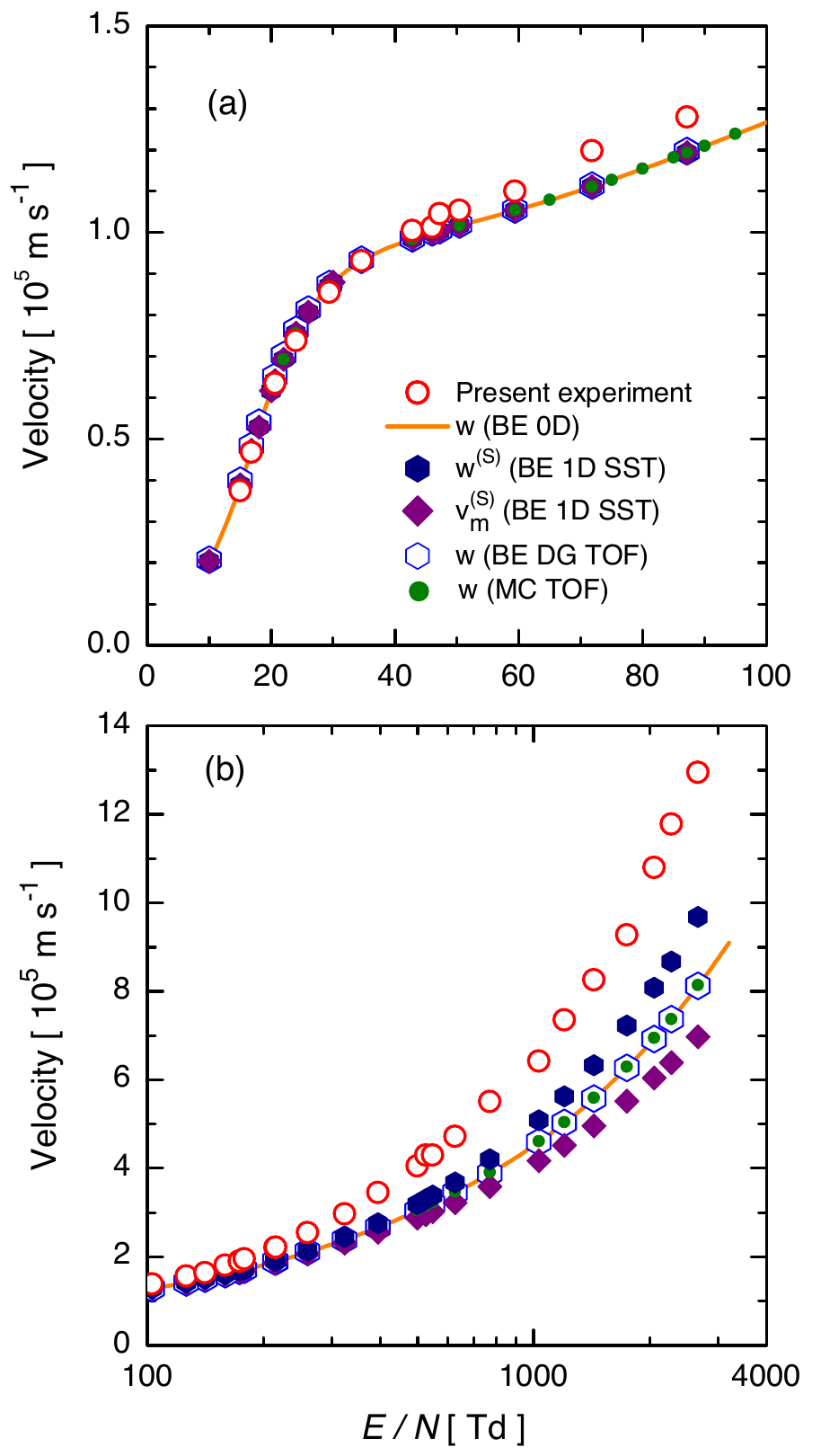} 
\caption{Flux drift velocity $w$ obtained from the different computational approaches.  The SST 
mean velocity $v_{\mathrm{m}}^\mathrm{(S)}$ and 
the present experimental data for the {\it bulk} drift velocity $W$ are shown for comparison. Panel (a) shows the low $E/N$ domain, and  panel (b) presents data for the higher range of $E/N$. The legend is the same for both panels.}
\label{fig:fluxv}
\end{center}
\end{figure}

\begin{figure}[!ht]
\begin{center}
\includegraphics[width=0.55\textwidth]{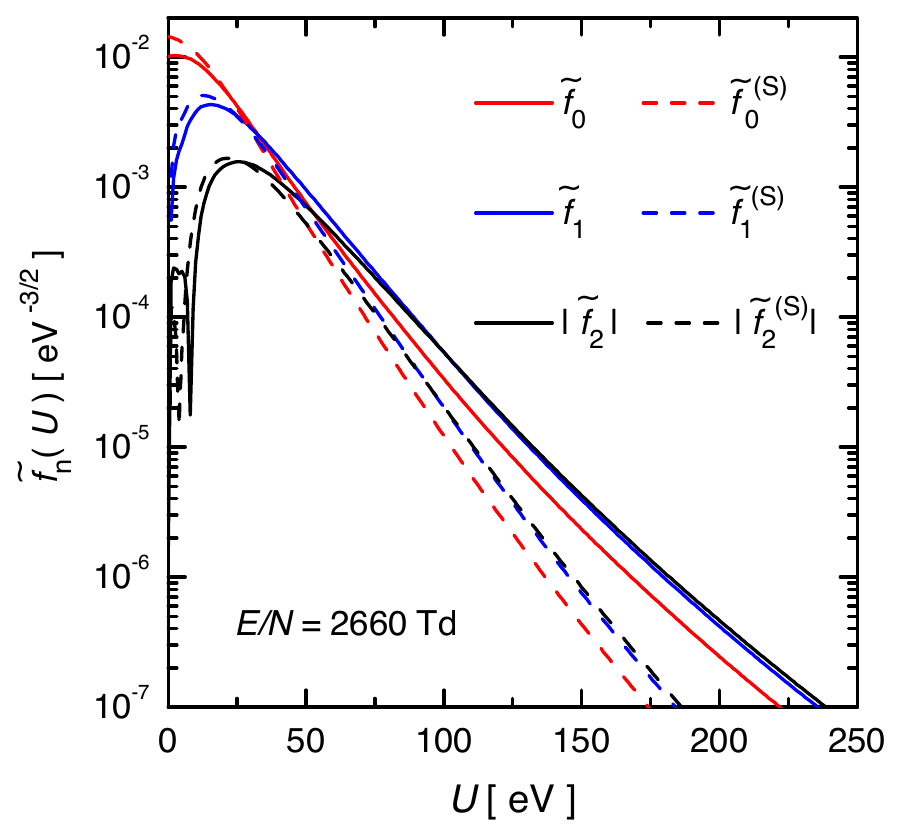} 
\caption{Expansion coefficients $\tilde{f}_n$  with $n=0-2$ 
obtained by method BE~0D 
as well as corresponding coefficients $\tilde{f}_n^\mathrm{(S)}$  obtained by method BE~1D~SST
for electrons in CO$_2$ at $E/N=\unit[2660]{Td}$. }
\label{fig:expansioncoefficients}
\end{center}
\end{figure}

Figure~\ref{fig:fluxv} displays the flux drift velocity $w$ obtained by the different methods 
as well as the mean velocity $v_\mathrm{m}^\mathrm{(S)}$ at SST conditions 
derived according to~(\ref{eq:1dmeanvelocity}). The latter is also referred to as diffusion-modified drift velocity in~\cite{TaSaSa1977JPDAP1051}. Furthermore, the present experimental data for the {\it bulk} drift velocity $W$ are shown for comparison. At low reduced electric fields (figure~\ref{fig:fluxv}(a)) all computational approaches give consistent results, 
indicating as well that the contribution of the electron attachment e.g.\ on $v_\mathrm{m}^\mathrm{(S)}$  is rather small.  However, at high $E/N$ (figure~\ref{fig:fluxv}(b)) the agreement is retained only for the methods BE~0D, BE~DG~TOF and MC~TOF. 
Because electron impact ionization processes are increasingly involved at larger $E/N$, 
$v_\mathrm{m}^\mathrm{(S)}$ is known to become less than the flux drift velocity~$w$~\cite{DuWhPe2008JPDAP245205}. At the same time the flux drift velocity~$w^\mathrm{(S)}$ at SST conditions assumes larger values than the flux drift velocity~$w$ 
obtained for the hydrodynamic regime of the time-dependent electron swarm. This finding is in agreement e.g.\ with the studies for argon reported in~\cite{TaSaSa1977JPDAP1051,BlFl1984AJP593}. 
It is also an immediate consequence of the establishment of different expansion 
coefficients $\tilde{f}_n$ (method BE~0D) 
and $\tilde{f}_n^\mathrm{(S)}$ (method BE~1D~SST), 
and thus the electron velocity distribution functions, under 
hydrodynamic and SST conditions, respectively. 
As an example, the corresponding first three expansion coefficients 
($n=0-2$) are shown in figure~\ref{fig:expansioncoefficients}. 
Similar results were also discussed  
 e.g.\ for synthetic air at larger $E/N$ values 
 in~\cite{HoLoVoBeBr2016PSST025017}. 
Moreover, the flux drift velocities shown in figure~\ref{fig:fluxv} 
remain 
increasingly smaller than the measured bulk drift velocity  $W$ 
at larger $E/N$ because of the additional impact of the effective 
ionization on the latter.

\begin{figure}[!ht]
\begin{center}
\includegraphics[width=0.55\textwidth]{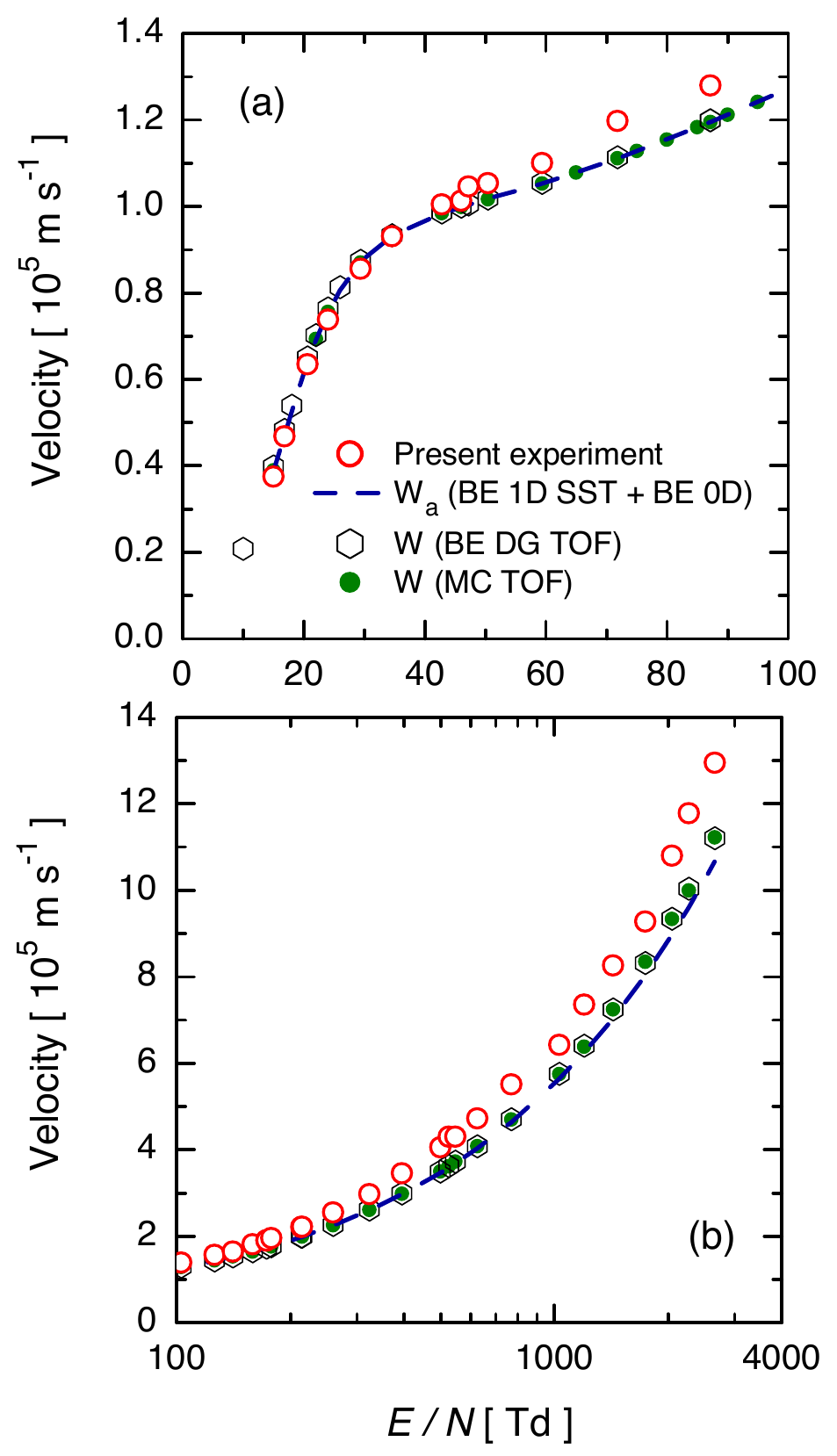} 
\caption{Bulk drift velocity $W$ 
obtained from the different computational approaches 
and comparison with present experimental data. 
Panel (a) shows the low $E/N$ domain, and panel (b) presents data for the higher range of $E/N$. The legend is the same for both panels.}
\label{fig:bulkv}
\end{center}
\end{figure}

Figure \ref{fig:bulkv} displays the bulk drift velocity $W$ computed by the MC TOF and DG TOF methods,  an approximate value $W_{\rm a}$ derived according to (\ref{eq:WfromSSTand0dresults}) 
from the combination of the calculated results of BE~1D~SST and BE~0D 
and the present experimental results. 
The values from the MC TOF and DG TOF methods agree perfectly, while $W_{\rm a}$ can be considered as a good approximation of $W$ up to about $ \unit[1000]{Td}$. The computed values are in good agreement with the experimental data at low fields of $E/N \lesssim \unit[50]{Td}$.  Above this field strength the measured values are consistently higher than the computational results. 
The deviations between the experimental data and the TOF results amount between 10 and \unit[15]{\%} above \unit[200]{Td}. 

To interpret these results we must keep in mind that the cross section set used~\cite{Grofulovic16} was developed using a two-term expansion code, BOLSIG+ \cite{0963-0252-14-4-011}. The transport parameters computed by this code can optionally include the effect of non-conservative processes and of the electron density gradients, depending on the problem being studied. 
In the derivation of the present cross section set the flux drift velocity \cite[figure 3]{Grofulovic16} was fitted at high $E/N$ to  Schlumbohm's results~\cite{SchlumbohmDrift} for the bulk drift velocity $W$. 
These results are lower than our and other experimental results at high 
$E/N$, as can be seen in figure~\ref{fig:exp-and-comp}. 
Thus, the present computational results for the flux drift velocity $w$ 
in fact reproduce the  experimental results of Schlumbohm and the apparent good fit of the computed values for the bulk drift velocity to the present experimental results in figure~\ref{fig:bulkv} is rather 
fortuitous. 


 \begin{figure}[!ht]
\begin{center}
\includegraphics[width=0.55\textwidth]{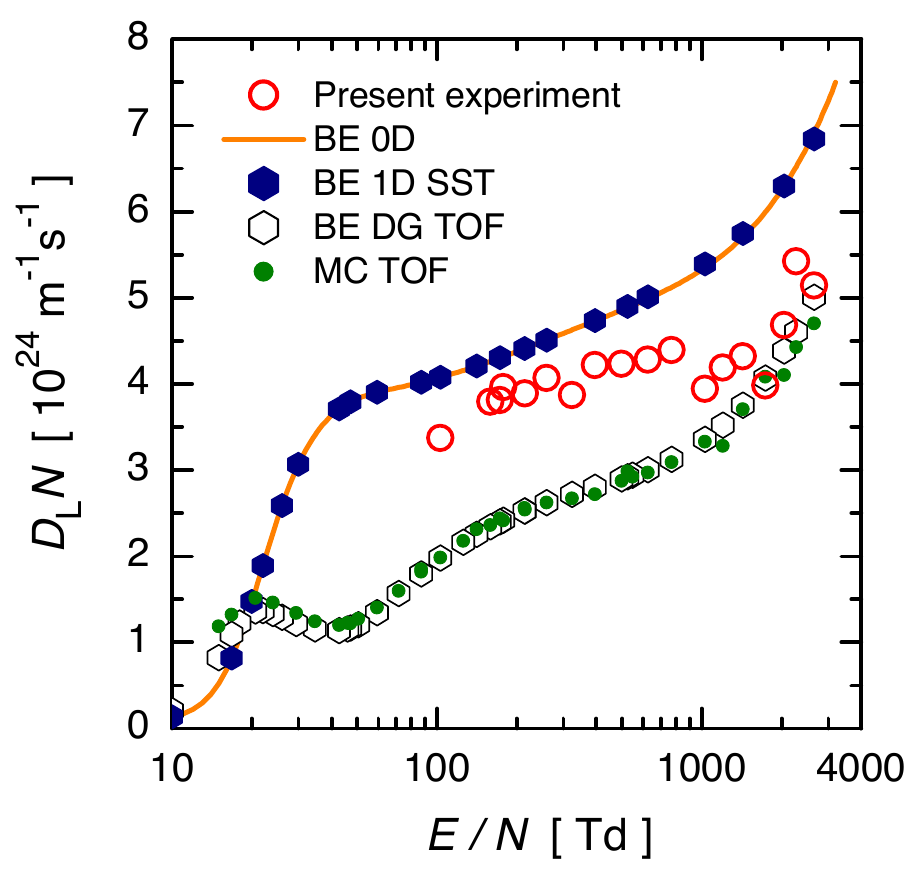} 
\caption{Computed values of the longitudinal diffusion coefficient, in comparison with the present experimental results.}
\label{fig:DL}
\end{center}
\end{figure}
 
\begin{figure}[!ht]
\begin{center}
\includegraphics[width=0.55\textwidth]{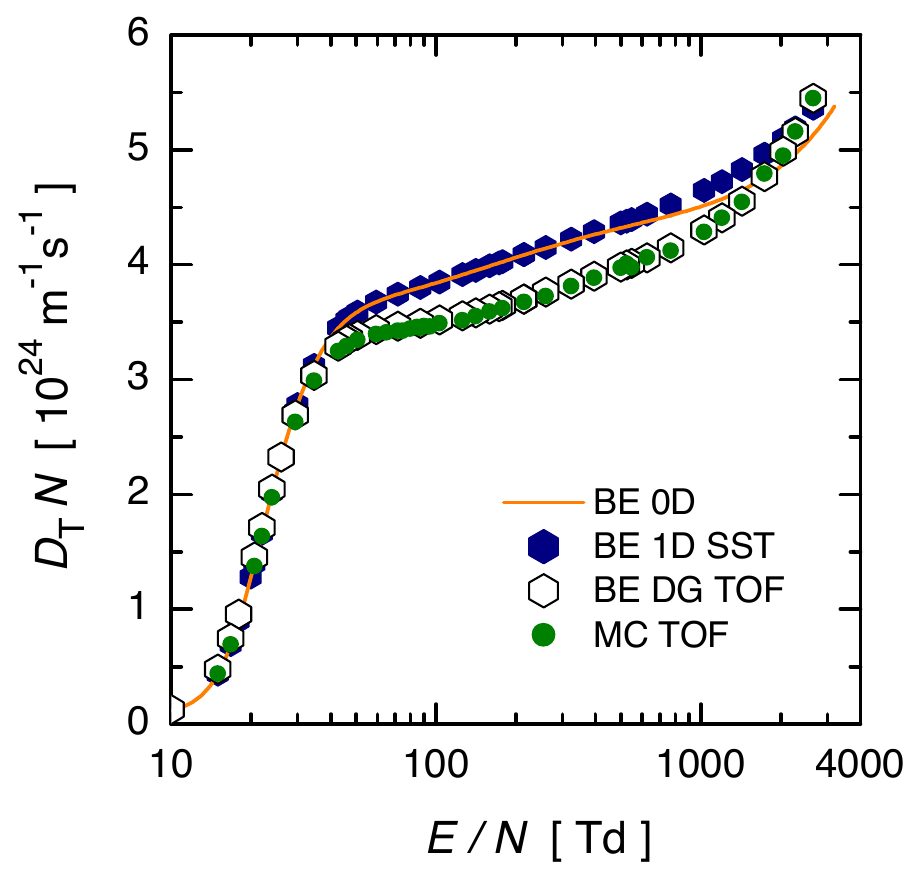} 
\caption{Computed values of the transverse diffusion coefficient.}
\label{fig:DT}
\end{center}
\end{figure}

Computed values of the longitudinal ($D_{\rm L}$) and transverse ($D_{\rm T}$) diffusion coefficients are shown in figures \ref{fig:DL} and \ref{fig:DT}, respectively. The values have been obtained by the BE~0D and BE~1D~SST solutions as well as by the BE~DG~TOF and MC~TOF methods. Available measured data for $D_{\rm L}$ are also shown in figure~\ref{fig:DL} for comparison. 
The results of the methods BE~0D and BE~1D~SST 
correspond to the respective flux components of the diffusion tensor and they 
practically overlap. The same is observed for the results 
obtained by the methods BE~DG~TOF and MC~TOF, which 
correspond to the bulk components of the diffusion tensor. 
Regarding the longitudinal diffusion 
coefficient (figure~\ref{fig:DL}), the 
BE~DG~TOF and MC~TOF results can directly be compared with the experimental results for $\DL$.   
None of the results, however, fits these experimental results for $\DL$.  In order to understand 
this finding, 
we note that in  \cite[figure~5]{Grofulovic16} the high $E/N$ values for the flux component of $D_L N$ 
are said to be 
fitted to the experimental results of  Schlumbohm~\cite{SchlumbohmDL},  which are below the present 
computed 
ones~(cf.~figure \ref{fig:exp-and-comp}). 
Rather, it seems that the 
 fit in~\cite[figure~5]{Grofulovic16} 
 was done 
using an expression equivalent to the first two terms of equation (\ref{eq:DG_DL2}), i.e., including the effect of the electron density expansion but neglecting the contribution from non-conservative 
processes~\cite[page~16]{Bolsig16}. This 
was found to be comparable with our 
results obtained by the BE DG TOF and MC TOF methods, if  we neglect the effect of non-conservative processes in the latter, 
i.e., assume $\nyeff=0$ and 
$\int \tnyeff(v) F_{zz}^{(2)}(\vec{v}) 
\mathrm{d}\vec{v} =0$
in~(\ref{eq:DG_DL2}).

\begin{figure}[!ht]
\begin{center}
\includegraphics[width=0.55\textwidth]{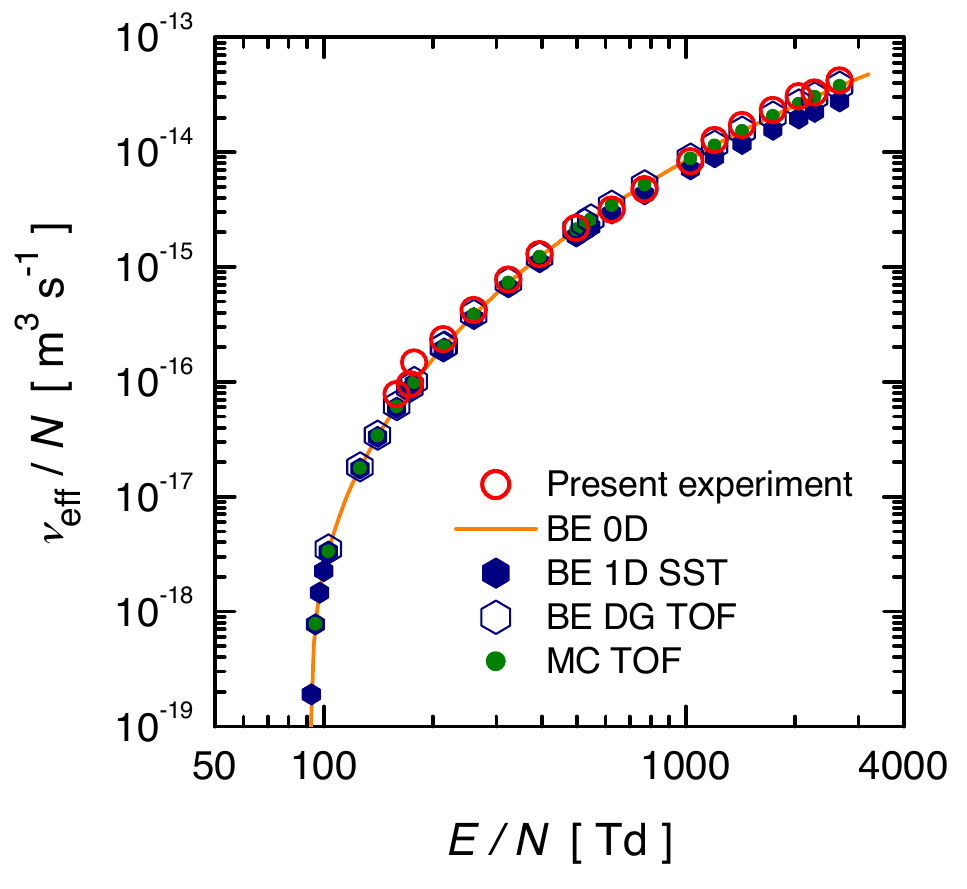} 
\caption{Reduced effective ionization frequency $\nu_{\rm eff}/N$ as a function of $E/N$.}
\label{fig:nu}
\end{center}
\end{figure}

The 
reduced ionization frequency $\nu_{\rm eff}/N$ 
 computed by the different approaches 
 as a function of the reduced electric field 
 is displayed in 
 figure~\ref{fig:nu}. 
 Good agreement between the measured and calculated results 
 is generally found. 
 This also holds for the reduced ionization frequency 
 $\nyeff/N$ 
 obtained by the BE~0D method, which perfectly agrees with 
 the swarm-averaged $\nyeff/N$ computed by the methods 
 BE~DG~TOF and MC~TOF. 
 Slight differences become visible 
 at larger $E/N$ only for the reduced ionization frequency 
 $\nyeff^{\mathrm{(S)}}/N$ obtained for 
 SST conditions using the method BE~1D~SST. This finding indicates  
 once more the differences between results obtained for 
 SST conditions and for the hydrodynamic regime 
 of time-dependent electron swarm studies. 
 Similar differences were reported e.g.\ for argon in~\cite{TaSaSa1977JPDAP1051,BlFl1984AJP593}. 
 
\begin{figure}[!ht]
\begin{center}
\includegraphics[width=0.45\textwidth]{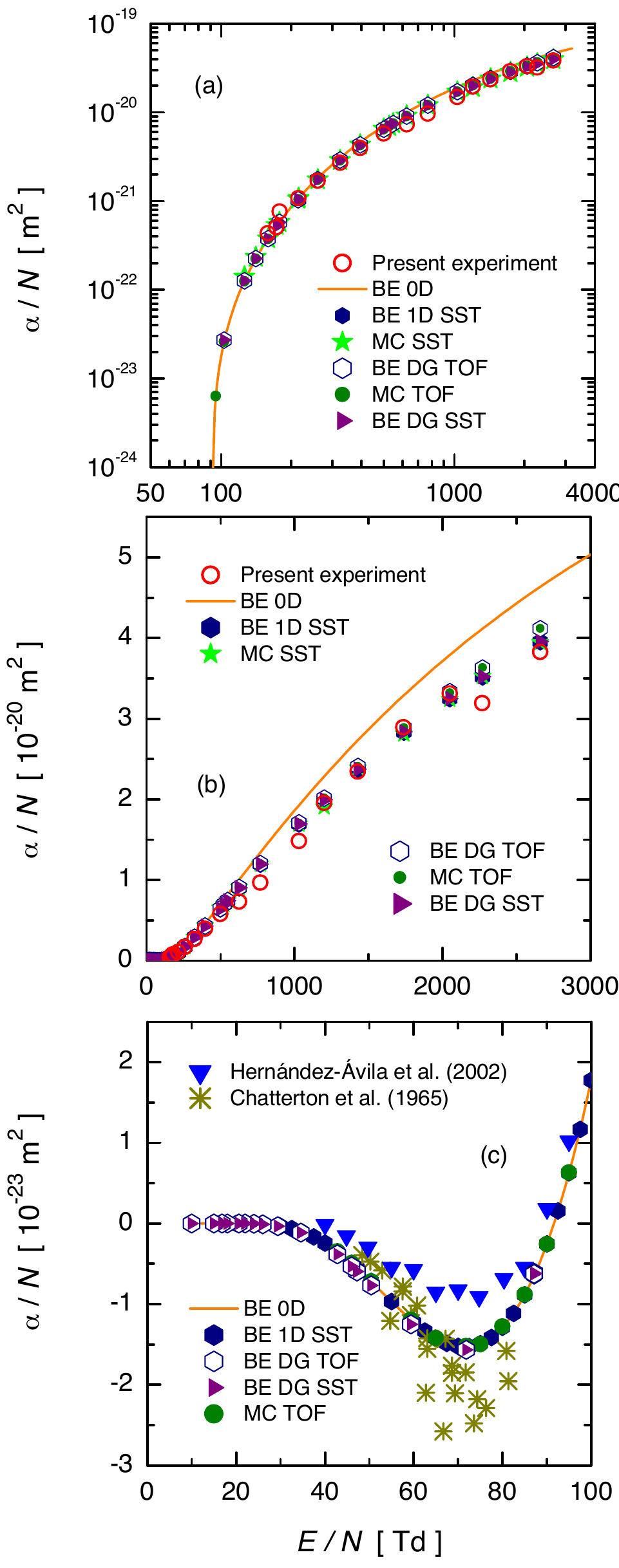} 
\caption{Reduced effective Townsend ionization coefficient  $\alpha/N$ as a function of $E/N$.  (a,b) Results obtained by the different computational methods in comparison with the present experimental data 
and $\alpha^{(0)}/N$ computed by method BE~0D.  (c) Low-field, attachment-dominated regime: comparison of the data obtained from the solution of the Boltzmann equation and MC simulation, as well as from previous experiments: Hern\'andez-\'Avila {\it et al.}~\cite{Hernandez} and Chatterton~{\it et al.} \cite{Chatterton}. 
}
\label{fig:alpha}
\end{center}
\end{figure}

Figure \ref{fig:alpha} shows the 
effective Townsend ionization coefficient $\alpha$ 
computed by the different SST and TOF methods (cf. table~\ref{tab:methods}) 
and the present 
experimental $\alpha$ obtained according to eq.~(\ref{eq:conn}) using the measured values of $W$, $\DL$ and $\nyeff$. Furthermore, 
the effective ionization coefficient $\alpha^{(0)}$ determined 
by the method BE~0D is displayed. 
Figure~\ref{fig:alpha}(a) 
 shows good agreement between all data sets 
 in the double logarithmic representation. 
Certain  differences 
 between the results calculated by the different 
methods 
 become obvious from the 
 linear representation in figure~\ref{fig:alpha}(b), while the 
 present 
 experimental data still agree quite well with the 
 computed $\alpha$. 

The consistent way to determine the effective Townsend ionization coefficient $\alpha$, relevant to SST conditions, is the use of  the 
methods BE~1D~SST, BE~DG~SST or MC~SST. The corresponding 
data for  $\alpha$ are in excellent agreement in 
figure~\ref{fig:alpha}(b).  Increasing differences  
from these results  for $E/N \geq \unit[500]{Td}$ are found 
for the effective ionization coefficient $\alpha^{(0)}$ 
(\ref{eq:alphafrom0D}) obtained by the BE~0D method. 
These differences are 
natural, because $\alpha^{(0)}$ and $\alpha$ are 
different coefficients. 

In addition, smaller, but increasing deviations between the 
results for $\alpha$ obtained by the TOF methods BE~DG~TOF and MC~TOF and 
those of the SST methods emerge above $E/N = \unit[2000]{Td}$. Because both the BE~DG~TOF and MC~TOF results agree excellently, it seems that the differences between the 
SST and TOF data result from the neglect of higher-order terms in the derivation of equations~(\ref{eq:conn}) and 
(\ref{eq:alphaTfromDGM}), respectively, used to determine 
$\alpha$ 
in the TOF case~\cite{TaSaSa1977JPDAP1051,BlFl1984AJP593}. 

The low-field domain ($E/N \leq$ 100 Td) is analyzed in figure~\ref{fig:alpha}(c) 
in more detail.  In this domain attachment dominates 
and ionization is hardly present.  The computational results are consistent and show reasonable agreement with the two experimental data sets of Hern\'andez-\'Avila {\it et al.} \cite{Hernandez} and Chatterton {\it et al.} \cite{Chatterton},  which are also shown in this figure.
 
 \begin{figure}[!ht]
\begin{center}
\includegraphics[width=0.45\textwidth]{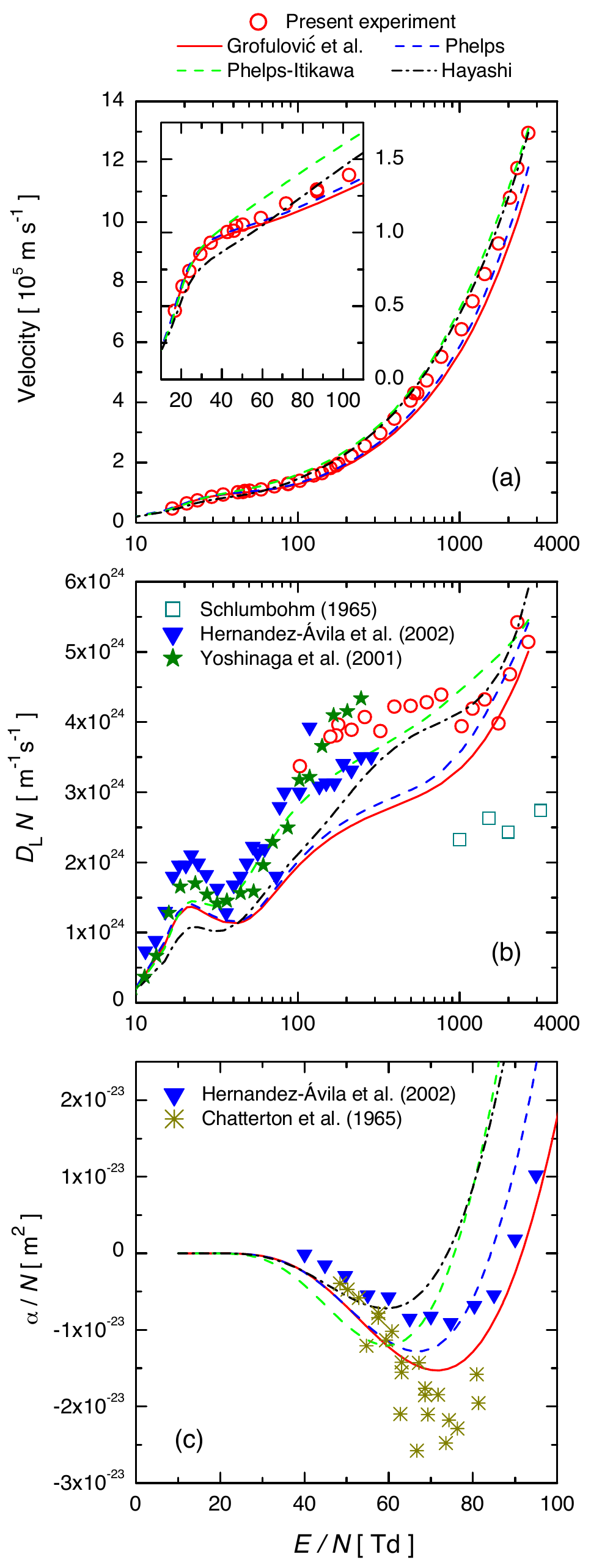} 
\caption{Measured values of the transport coefficients in comparison with  values computed by method BE~DG~TOF (a,b) and 
BE~1D~SST 
(c) using different cross section sets: Grofulovi\'c {\it et al.} \cite{Grofulovic16}, Phelps \cite{PhelpsCS}, Phelps-Itikawa (see text) \cite{PhelpsCS,Itikawa02}, and Hayashi \cite{Hayashi90}. The experimental data are repeated from previous figures, for references see the captions of figures \ref{fig:exp-and-comp} and \ref{fig:alpha}.}
\label{fig:12}
\end{center}
\end{figure}

\section{Concluding remarks}
\label{sec:Summary}

We have investigated electron transport in \CO~gas experimentally using a scanning drift tube, as well as computationally by
solutions of the electron Boltzmann equation and via Monte Carlo simulation, corresponding to both time-of-flight and steady-state Townsend conditions.
The experimental system operated under TOF conditions and allowed recording the spatio-temporal evolution of electron swarms initiated by short UV laser pulses. The measured data made it possible to derive the bulk drift velocity, the longitudinal diffusion coefficient, and the effective ionization frequency of the electrons, for the wide range of the reduced electric field 
from 15 to $\unit[2660]{Td}$. 
The measured TOF transport coefficients and the effective Townsend ionization coefficient, deduced from these coefficients, have been compared to experimental data obtained in previous   
studies, where generally good consistency with most of the transport parameters  obtained in these earlier studies 
 was found. 

Comparison of the experimental data was 
also 
carried out with transport coefficients resulting from 
various 
kinetic computations, which used a cross section set published recently~\cite{Grofulovic16}. 
The computational results point out the range of applicability 
 of the respective methods used to determine the 
 different measured transport properties of electrons in 
 CO$_2$. In particular, 
significant 
differences between 
our measured and  computed (by the TOF methods) 
values of the bulk drift velocity and the longitudinal diffusion coefficients have been found, which are 
attributable 
to the specific methods involved in the construction of the cross section set.
Namely, this cross section set was developed neglecting the effect of non-conservative processes and fitting, at high $E/N$, the results of Schlumbohm~\cite{SchlumbohmDrift,SchlumbohmDL} for $W$ and $\DL N$, which are significantly different from the present experimental results. 
Thus, 
the experimental results for the bulk drift velocity and longitudinal diffusion coefficients 
could not 
be 
reproduced correctly. 
However,  the computational results obtained by all the methods used in this paper to determine the transport properties in the hydrodynamic regime  
show 
good agreement for the effective ionization frequency.  Furthermore, the effective Townsend ionization coefficient calculated by the SST and TOF methods agree well with the results obtained from the experiments for the large range of $E/N$ investigated.  

Other cross section sets, which are available for \CO~gas,  do not perform better 
as well 
for all conditions and transport coefficients. 
A comparison of the computed bulk drift velocity, longitudinal diffusion coefficient and 
effective Townsend 
ionization coefficient using several different, widely used sets of cross sections is presented in figure~\ref{fig:12}. The cross section sets include the one by 
Grofulovi\'c 
\textit{et al.}~\cite{Grofulovic16}, 
that of 
Phelps~\cite{PhelpsCS}, 
a modified cross section set of  Phelps using the elastic momentum transfer cross section of Itikawa~\cite{Itikawa02} (named Phelps-Itikawa set in the following), as well as the set recommended by Hayashi~\cite{Hayashi90}. Figure \ref{fig:12}(a) reveals that the drift velocity $W$ is best reproduced by the Phelps and 
the  
Grofulovi\'c 
\textit{et al.} 
cross sections at low $E/N$, while none of the computed values fits well our  
experimental data points for the $200 \leq E/N \leq 
\unit[1500]{Td}$ range. A better agreement for the last few (high $E/N$) data points is found for the Phelps-Itikawa set and the Hayashi set. 
Computed values of $\DL$ 
(figure~\ref{fig:12}(b)) 
follow most closely the measured data up to moderate reduced electric field values ($E/N \lesssim \unit[1000]{Td}$) when the Phelps-Itikawa cross section 
set 
is used. At higher $E/N$ values the only experimental data 
in addition to Schlumbohm's 
results~\cite{SchlumbohmDL} 
originate from the present measurements. 
Our data 
show a larger scattering, while the computed curves become closer 
and incidentally agree  quite well with them. 
Regarding the 
effective ionization 
coefficient (figure \ref{fig:12}(c)) 
no precise agreement is found for the low 
(attachment-dominated) 
$E/N$ domain with any of the computational results obtained with the different cross section sets.

Thus, we think that our present experimental results offer large potential for future adjustments of electron collision cross sections for CO$_2$.

\ack This work was supported by the Hungarian Fund for Scientific Research (OTKA), via grant K105476 and Funda\c{c}\~{a}o para a Ci\^{e}ncia e a Tecnologia (FCT, Portugal), under projects UID/FIS/50010/2013 and UID/FIS/PTDC/FIS-PLA/1420/2014 (PREMiERE). We gratefully acknowledge the contributions of T. Sz\H{u}cs and P. Hartmann to the development of the experimental apparatus. 
The studies were performed in the framework of the Collaborative Research Centre Transregio 24 ``Fundamentals of Complex Plasmas''.

\section*{References}

\bibliographystyle{unsrt}
\bibliography{references}

\end{document}